\documentstyle[prc,aps]{revtex}

\begin{document}
\draft                                                            
\twocolumn

\title{Neutrinoless Double Beta Decay within 
QRPA with Proton-Neutron Pairing}

\author{ G. Pantis$^1$\thanks{{\it e-mail:} gpantis@cc.uoi.gr},
F. \v Simkovic$^2$\thanks{{\it On leave from:} 
Bogoliubov Theoretical Laboratory, 
Joint Institute for Nuclear Research, 
141980 Dubna, Moscow Region, Russia and Department of Nuclear Physics,  
Comenius University, Mlynsk\'a dolina F1, Bratislava, Slovakia},
J.D. Vergados$^1$ and Amand Faessler$^2$}
\address{1.  Theoretical Physics Section, University of Ioannina,\\
 GR 451 10, Ioannina, Greece \\
 2.  Institute f\"ur Theoretische Physik der Universit\"at 
T\"ubingen\\ 
 Auf der Morgenstelle 14, D-7a076 T\"ubingen, Germany }
\date{\today}
\maketitle
\begin{abstract}
We have investigated the role of proton-neutron pairing in the  context
of the Quasiparticle Random Phase approximation formalism. This way
the neutrinoless double                           
beta decay matrix elements of the experimentally interesting 
A= 48, 76, 82, 96, 100, 116, 128, 130 and 136 systems have been 
calculated. We have found that the inclusion of proton-neutron pairing
influences the neutrinoless double beta decay rates significantly,
in all cases allowing for larger values of the expectation value of light
neutrino masses.
Using the best presently available experimental limits on the 
half life-time of neutrinoless double beta decay we have extracted the
limits on lepton number violating parameters. 
\end{abstract}
\pacs{23.40.Hc}

\narrowtext

\section{ Introduction}

Among the exotic processes the neutrinoless double $\beta$-decay 
($0\nu \beta\beta$-decay)
  
\begin{eqnarray}
(A,Z) \rightarrow (A,Z \pm 2) + e^{\mp} +  e^{\mp}, 
\label{eq:1}  
\end{eqnarray}
has been sought experimentally for about half a century but it has not yet
been observed \cite{1}.  Its observation will undoubtedly be a signal of
interesting physics beyond the standard model  of
electroweak interactions.  First of all it will demonstrate the breakdown
of lepton number conservation which, being associated with a global, not
gauge, symmetry is expected to be broken at some level. It will also give
us useful information about the neutrino mass if it is in the region $\leq
1$ eV.  Finally, and most importantly, it is the best process, if not the
only one, to settle the issue of whether the neutrino mass eigenstates are
of the Majorana type, i.e.,  whether the particle coincides with its own
antiparticle (a la $\pi^0$), or of the Dirac type (a la $K^0$).

It is, therefore, not surprising  that the experimental searches 
\cite{2}-\cite{15} for
the above process have persisted with great devotion up to the present day
and have lead to the unbelievably long life time limit \cite{2}
\begin{center}
$T^{0\nu}_{1/2} > 5.6 \times 10^{24} y $.  
\end{center}
From this limit, in conjunction with calculations of the nuclear
matrix elements involved, the limit $|<m_{\nu}>| \leq 0.8$ eV has been
extracted for the average light neutrino mass.  It is interesting to remark
that neutrino masses in this neighborhood can constitute candidates for the
Hot Dark Matter (HDM) component required for the understanding of the
large scale  structure of the universe as indicated by the KOBE data
\cite{16,17}.

The analysis of the $0\nu \beta\beta$-decay data, if and when they become
available, is unfortunately not going to be simple.  Depending upon the
extension of the standard model assumed there are many mechanisms which can
lead  to process (\ref{eq:1}) which can interfere with one another (see refs.
\cite{18}-\cite{22} for reviews).  
In those mechanisms in which the exchanged particle is light (e.g., light
Majorana neutrino) the effective transition operator is of long range 
(essentially coulombic) with or without spin dependence.  If on the other
hand the exchanged particle is heavy (e.g., heavy Majorana neutrino, exotic
Higgs Scalars etc.) the effective operator is of a short range and in the
presence of short range correlations, one must be careful not to ignore
\cite{18} the finite size of the nucleon ($\sim0.8$ fm).  In this case the size
of the operator is set by the size of the nucleon.  It is obvious that the
two cases involve different nuclear physics and as we shall see later
they lead to the extraction of different parameters of the gauge theory.

It is clear from the above discussion that all nuclear matrix elements must
be computed reliably. This is easier said than done, however, since the
nuclear sytems which can undergo double $\beta$-decay, with the  possible
exception of the $A=48$ system, are far away  from closed shells  and as
a result have complicated structure \cite{18}-\cite{21}.  In the shell
model approach it is clearly impossible to construct all the
needed intermediate states of the nucleus which is one unit of change away. 
Since, however, the energy denominators are dominated by the virtual
neutrino momentum, rather than the nuclear excitation energy, the 
construction of all these states can be avoided using a version of the
closure  approximation \cite{18}-\cite{21}.  But even then it is quite hard  to
construct the wave functions of the initial and final nuclei employing a
full shell model basis.   Thus a weak coupling scheme has been employed,
starting from products of neutron and proton wave  functions and employing
truncations according to the energies of the unperturbed proton and neutron
states.  In any case it is quite clear that it is very hard to substantially
improve these old calculations \cite{23}-\cite{25}.

It is thus not surprising that other approximation schemes have been
employed. The most prominent among them has been the Quasiparticle Random 
Phase Approximation \cite{26}-\cite{34} (QRPA).  
In this scheme the construction of the
intermediate states is unavoidable, but not extremely hard.  In the first
step one constructs the intermediate states $|mJM>$ as proton-particle
neutron - hole excitations built on the ground state of initial nucleus.  In
the second step one views the intermediate states $|{\overline m}JM>$ as
neutron-particle proton-hole, built on the ground state of the final
nucleus.  These intermediate states are expressed as two quasiparticle
states and one makes proper adjustments  for the fact that the two sets of
states $|m J_{\text{m}}>$  and 
$|{\overline m} J_{\text{m}}>$ are not orthogonal.

In the QRPA approximation there is no need to invoke the closure
approximation \cite{26}-\cite{28}.  
In  fact it was  possible to use QRPA to explicitly
check how well the closure approximation works.  It was rewarding  to find
that it works quite well except in those situations when the matrix
elements are unusually suppressed.

As we have previously mentioned a number of QRPA calculations 
\cite{24}-\cite{38} for almost all nuclei of practical 
interest in $0\nu \beta\beta$-decay have been performed.  In all such
calculations the two quasiparticle states were of the proton-neutron
 variety.  Proton-neutron (p-n) pairing correlations \cite{39}-\cite{43}
 had been neglected. Such
correlations were recently found, however, to be important 
\cite{44,45,46} in the
evaluation of the nuclear matrix elements entering $2\nu \beta\beta$-decay
\begin{equation}
(A,Z)  \rightarrow  
(A,Z + 2) + e^- + e^-  + {\tilde \nu}_{\text{e}}  +  
{\tilde \nu}_{\text{e}}, 
\label{eq:2}  
\end{equation}
\begin{equation}
(A,Z)  \rightarrow  (A,Z - 2) + e^+ + e^+ + \nu_{\text{e}}  
+ \nu_{\text{e}},  
\label{eq:3}  
\end{equation}
which proceeds only via the $1^+$  intermediate nuclear states.

It is the purpose of the present paper to investigate what effect, if any,
the p-n pairing has on the $0\nu \beta\beta$-decay matrix
elements.  To this end we will repeat and extend our previous
calculations 
\cite{26,27,28} to cover most of the nuclear targets of experimental
interest
($^{48}{\text{Ca}}$, $^{76}{\text{Ge}}$, $^{82}{\text{Se}}$, 
$^{96}{\text{Zr}}$, $^{100}{\text{Mo}}$,
$^{116}{\text{Cd}}$, $^{128}{\text{Te}}$, $^{130}{\text{Te}}$, 
$^{136}{\text{Xe}}$).

\section{  Lepton violating parameters and associated nuclear matrix
elements}

As we have mentioned in the introduction there exist many mechanisms which
can lead to  $0\nu  \beta\beta$ decay some of which are exotic (exotic Higgs
scalars,  supersymmetric R-parity violating interactions etc. 
\cite{18}).
The most  popular scenario is
one which involves intermediate massive Majorana neutrinos.  We will
concentrate on the last mechanism in this work even though some of the
nuclear matrix elements computed may be used for the more exotic
mechanisms as well.

We will distinguish the following two cases:

i) Both leptonic currents are of  the same chirality, i.e., both 
left handed or both right handed.  Then out of the neutrino propagator one
picks the part
\begin{equation} 
\frac {m_{\text{j}}} {q^2 - m^2_{\text{j}}},  
\label{eq:4}  
\end{equation}
which for light neutrinos yields an amplitude proportional to
 \begin{equation}
<m_\nu> = \sum_{\text{j}} |U^{(11)}_{\text{ej}}|^2 
m_{\text{j}} e^{ {\text{i}}\alpha_{\text{j}} }, 
\label{eq:5}   
\end{equation}
 $\exp(\text{i}\alpha_{\text{j}})$  is the CP eigenvalue of the neutrino
mass  eigenstate $|\nu_{\text{j}}>$. 
For light neutrinos the contribution of the
right handed currents is negligible.  For heavy neutrinos the amplitude
becomes proportional to
\begin{equation}
<\frac{1}{M_{\text{N}}}>^{}_{\text{L}} 
= \sum_{\text{j}} |U^{(12)}_{{\text{ej}}}|^2 \frac{1}{M_{\text{j}}} 
e^{{\text{i}}\phi_{\text{j}}},
\label{eq:6}
\end{equation}
\begin{equation}
<\frac{1}{M_{\text{N}}}>^{}_{\text{R}} 
= \sum_{\text{j}} |U^{(22)}_{{\text{ej}}}|^2 \frac{1}{M_{\text{j}}}
e^{{\text{i}}\phi_{\text{j}}},
\label{eq:7}
\end{equation}
for left and right handed currents respectively.  
$\exp (\text{i}\phi_{\text{j}})$ and $M_{\text{j}}$
are the CP eigenvalue and the mass of  the heavy neutrino mass
eigenstate $|N_{\text{j}}>$.  
The sub-matrices $U^{(11)}$, $U^{(12)}$, $U^{(21)}$,
$U^{(22)}$ are the parts of the unitary matrix which connect the weak
eigenstates 
$(\nu^0_\alpha, \nu^{0{\text{c}}}_\alpha), \alpha =e, \mu, \tau,$ with
the mass eigenstates $(\nu_{\text{i}}, N_{\text{i}}),$ 
i = 1, 2, 3 (see e.g.,  ref.  \cite{18}).
The upper index ``1'' stands for the light and ``2'' for the heavy
neutrinos of left-right symmetric models.  For each
case one encounters a spin independent operator 
($\text{vector}\times\text{vector}$ or Fermi)
and one which is a spin dependent scalar 
($\text{axial}\times\text{axial}$ or Gamow-Teller), 
yielding the matrix elements $M_{\text{F}}$ and 
$M_{{\text{GT}}}$ respectively.
 
ii)
leptonic currents of opposite chirality.\\  
Then the relevant part of the
intermediate neutrino propagator is
\begin{equation}
\frac {q_\lambda} {q^2 - m^2_{\text{j}}}.  
\label{eq:8}  
\end{equation}
Since the right-handed currents are expected to be suppressed, 
this contribution is expected to be significant for light neutrinos. 
Thus the amplitude does not explicitly depend on the neutrino mass.  One can
now extract from the data two dimensionless parameters \cite{18}
\begin{equation}
\lambda  ~ \cong ~ \kappa ~\eta^{}_{{\text{RL}}},
\label{eq:9}  
\end{equation}
\begin{equation}
\eta ~  \cong  ~ \epsilon  ~ \eta^{}_{{\text{RL}}},
\label{eq:10}  
\end{equation}
where
\begin{equation}
\kappa ~ = ~ (M^{}_{{\text{L}}} / M_{{\text{R}}})^2,
\label{eq:11}  
\end{equation}
\begin{equation}
\epsilon ~ = ~ tg ~\zeta ~~({\text{mixing}}).
 \label{eq:12}
\end{equation}
$M_{{\text{L}}}$ and $M_{{\text{R}}}$ 
are respectively the masses of the vector bosons
$W^{}_{{\text{L}}}$ and $W^{}_{{\text{R}}}$
associated with  the left and right handed interactions. 
$\zeta $ is the $W^{}_{{\text{L}}}-W^{}_{{\text{R}}}$ mixing angle. 
$ \eta^{}_{{\text{RL}}}$ is 
given by \cite{18}
\begin{equation}
\eta^{}_{{\text{RL}}} = 
\sum_{\text{j}} U^{(11)}_{{\text{ej}}} U^{(21)}_{{\text{ej}}} 
e^{{\text{i}}\alpha_{\text{j}}}.
\label{eq:13}  
\end{equation}
The $\lambda$-term arises when the chiralities of the hadronic currents
match those of the leptonic currents, i.e.,  they  are of the  
$J_{\text{L}}-J_{\text{R}}$
combination.  The $\eta$-term arises  when the two hadronic currents are
of the same chirality, i.e of the  
$J_{\text{L}}-J_{\text{L}}$ type (this is possible due to
the W-boson mixing).  For the extraction of $\lambda$
 one
should know five matrix elements i.e., ($M_{{\text{F}}\omega},~
M_{{\text{GT}}\omega}$) 
which arise from the time component of the propagator of 
Eq.\ (\ref{eq:8}) and ($M^{\prime}_{\text{F}}, 
M^{\prime}_{{\text{GT}}}, M^{\prime}_{\text{T}}$ (tensor))
arising from the space component of that propagator.  Due to the
different energy dependence $M_{{\text{F}}\omega}, 
M_{{\text{GT}}\omega}, M^{\prime}_{\text{F}}$ and
$M^{\prime}_{{\text{GT}}} $ are different from the 
two matrix elements $M_{\text{F}}$ and $M_{{\text{GT}}}$
encountered in the light neutrino mass mechanism. 
We will see, however,that to a
good approximation
\begin{equation}
M_{\text{F}} = M^\prime_{\text{F}} = M_{{\text{F}}\omega}, 
\label{eq:14}  
\end{equation}
\begin{equation}
M_{{\text{GT}}} = M^\prime_{{\text{GT}}} = M_{{\text{GT}}\omega}.
\label{eq:15}  
\end{equation} 
The situation is a bit more complicated in the case of 
the $\eta$-term since
one encounters two additional matrix elements $M^\prime_{\text{P}}$ 
and $M_{\text{R}}$.  The
first arises from an operator which is antisymmetric both  in the spin and
angular momentum indices \cite{18}-\cite{21}.
The second matrix element arises from the
momentum dependent terms of the hadronic current \cite{25}
(weak magnetism etc.). 
The contribution of the momentum dependent term is normally a small
correction.  This is not true in this case, however, since due to the
structure of the propagator 
of Eq.\ (\ref{eq:8}), the standard leading non-vanishing term is
proportional to the average lepton momenta \cite{1}.

One must note, furthermore, that the kinematics are different, 
reflecting the difference between (\ref{eq:4}) and (\ref{eq:8}).  Thus the
coefficients entering  the various combinations of nuclear matrix elements 
in the decay rate are
energy dependent.  As a result the relative importance of the various
nuclear matrix elements may vary from nucleus to nucleus (depending on the
available energy).

In the case of heavy neutrino intermediate states one encounters two 
matrix elements $M_{{\text{HF}}}$ and $M_{{\text{HGT}}}$ 
which differ from the above matrix
elements $M_{{\text{F}}}$ and $M_{{\text{GT}}}$ 
due to the fact that the radial part of the
relevant operator is short-ranged.

The nuclear matrix elements mentioned above are associated with a set of
transition operators which in momentum space can be cast in the general form
(see refs. \cite{26,27,28,39} for details)
\begin{equation}
\Omega = \sum_{\text{i}\neq \text{j}} \tau^+_{}(i) 
\tau_{}^+(j) \omega(i,j) g_{\text{m}} ({\bf q}_{\text{i}}, 
{\bf q}_{\text{j}}), 
\label{eq:16}  
\end{equation}
with
\begin{equation}
 g_{\text{m}} 
({\bf q}_{\text{i}}, {\bf q}_{\text{j}}) 
= \frac {4\pi R_0}{(2\pi)^3} \frac {1}{2\sqrt{2}}
\frac {\delta ({\bf q}_{\text{i}} + {\bf q}_{\text{j}})} 
{\Delta_{\text{m}} (q_{\text{i}}\sqrt{2})},
\label{eq:17}  
\end{equation}
\begin{equation}
R_0 = r_0 A^{1/3} \,\, (r_0 \simeq 1.1. \text{fm})
\ \ \ \ (\text{nuclear} \ \  \text{radius}),
\label{eq:18}  
\end{equation}
\begin{equation}
\Delta_{\text{m}} (q) 
= \sqrt{{ q}^2 +m^2_\nu}~ [\epsilon_{\text{m}} 
+\sqrt{{ q}^2 +m^2_\nu}].
\label{eq:19}  
\end{equation}
where $m_\nu$ is the mass of the virtual neutrino and 
$\epsilon_{\text{m}}$ a suitable energy
denominator [see Eq.\ (\ref{eq:46}) below].

The most important matrix element $M_{\text{GT}}$ is associated with
${\vec \sigma}_1 \cdot {\vec \sigma}_2,$ i.e.,
\begin{equation}
\omega_{\text{GT}}(1,2) 
=  {\vec \sigma}_1 \cdot {\vec \sigma}_2
 \longleftrightarrow M_{\text{GT}}. 
\label{eq:20}  
\end{equation}
Similarly
\begin{equation}
\omega_{\text{F}}(1,2) = 1 \longleftrightarrow M_{\text{F}}. 
\label{eq:21}  
\end{equation}

The matrix elements $M_{\text{FH}}, M_{\text{GTH}}$ 
associated with heavy neutrino are
related to the operators $\omega_{\text{FH}}$ and 
$ \omega_{\text{GTH}}$ where
\begin{equation}
\omega_{\text{FH}}(1,2) = \frac{m^2_\nu}{m_{\text{e}}m_{\text{p}}}
 \longleftrightarrow M_{\text{FH}},
\label{eq:22}  
\end{equation}
\begin{equation}
\omega_{\text{GTH}}(1,2) 
= \frac{m^2_\nu}{m_{\text{e}}m_{\text{p}}}
{\vec \sigma}_1 \cdot {\vec \sigma}_2 
\longleftrightarrow M_{\text{GTH}}.
\label{eq:23}   
\end{equation} 
Notice that for heavy neutrino
\begin{equation}
  \frac{m^2_\nu}{m_{\text{e}}m_{\text{p}}} 
\frac{1}{\Delta_{\text{m}}(q)} \approx
\frac{1}{m_{\text{e}} m_{\text{p}}} \ \  
\label{eq:24}   
\end{equation}
(independent of momentum)
to be compared with
\begin{equation}
\frac{1}{\Delta_{\text{m}} (q)} \approx
\frac{1}{ q^2} \ \  (\text{for} \ \text{light} \ \text{ neutrino}). 
\label{eq:25}   
\end{equation}

For processes which do not explicitly depend on the neutrino mass
($j_{\text{L}}-j_{\text{R}}$ interference) 
we encounter the operators $\omega_{\text{F}\omega}$ and
$\omega_{\text{GT}\omega}$ which differ from 
$\omega_{\text{F}}$ and $\omega_{\text{GT}}$ by the
inclusion of the extra kinematical factor 
$\delta_{\text{m}} (q_1 \sqrt{2})$ with
\begin{equation}
\delta_{\text{m}} (q) = \frac{  \sqrt{q^2 + m^2_\nu} }
{\epsilon_{\text{m}} + \sqrt{q^2 + m^2_\nu}}. 
 \label{eq:26}   
\end{equation}
One also encounters the matrix elements $M^\prime_{\text{F}}$ 
and $M^\prime_{\text{GT}}$
associated with the operators
\begin{equation}
\omega^\prime_{\text{F}} (1,2) = - 2 {\vec q}_1 \cdot {\vec \nabla}_1 
 \longleftrightarrow M^\prime_{\text{F}},
\label{eq:27}   
\end{equation}
\begin{equation}
\omega^\prime_{\text{GT}} (1,2) = - 2 
{\vec \sigma}_1 \cdot {\vec \sigma}_2 
\, {\vec q}_1 \cdot {\vec \nabla}_1
 \longleftrightarrow M^\prime_{\text{GT}}, 
\label{eq:28}   
\end{equation}
and the matrix element  $M^\prime_{\text{T}}$ associated with
\begin{eqnarray}
\lefteqn{ \omega^\prime_{\text{T}} (1,2) 
= \frac{1}{3} {\vec \sigma}_1 \cdot {\vec \sigma}_2 {\vec
q}_1 \cdot {\vec \nabla}_1 - {\vec \sigma}_1 \cdot {\vec q}_1 {\vec
\sigma}_2  \cdot {\vec \nabla}_1 }
&& \nonumber \\ 
&& \phantom{\omega^\prime_{\text{T}} (1,2) =}
- {\vec \sigma}_1 \cdot {\vec \nabla}_1
{\vec \sigma}_2 \cdot {\vec q}_1
 \longleftrightarrow M^\prime_{\text{T}}.   
\label{eq:29}   
\end{eqnarray}
As it has been mentioned above in the case of 
$j_{\text{L}}-j_{\text{R}}$ but $J_{\text{L}}-J_{\text{L}}$ 
combination we encounter two additional matrix elements  namely 
$M^\prime_{\text{P}}$ and $M_{\text{R}}$. 
$M_{\text{P}}^\prime$ is associated with the operator 
$\omega^{\prime}_{\text{P}} (1,2)$ :
\begin{equation}
\omega^{\prime}_{\text{P}} (1,2) = 
\frac{1}{2} ({\vec \sigma}_1 - {\vec \sigma}_2).  
({\bf \ell}_1 - {\bf \ell}_2) \longleftrightarrow 
M^{\prime}_{\text{P}}, 
\label{eq:30}   
\end{equation}
while $M_{\text{R}}$ is associated with 
the operator $\omega^{}_{\text{R}}$ :
\begin{equation}
\omega^{}_{\text{R}} \simeq 
\frac {g^{}_{\text{V}}}{g^{}_{\text{A}}} 
\frac {2\mu}{m_{\text{e}} m_{\text{p}} a^2} [\frac{2}{3}
\omega^{\text{R}}_{\text{S}} - \omega_{\text{T}}^{\text{R}}] 
\longleftrightarrow M_{\text{R}}, 
\label{eq:31}   
\end{equation}
with $\mu = \mu_{\text{p}} - \mu_{\text{n}+1} = 4.7$ 
(a is the oscillator length) and
\begin{equation}
\omega^{\text{R}}_{\text{S}} (1,2) = 
({\vec \sigma}_1 \cdot {\vec \sigma}_2) q^2_1 a^2,
\label{eq:32}   
\end{equation}
\begin{equation}
\omega^R_T (1,2) = ({\vec \sigma}_1 \cdot {\hat q}_1 \ \ 
{\vec \sigma}_2 \cdot {\hat q}_1 - \frac{1} {3}
{\vec \sigma}_1 \cdot {\vec \sigma}_2) q^2_1 a^2.
\label{eq:33}   
\end{equation}

In the expression for the $0\nu \beta \beta$-decay life time various
combinations of the above nuclear matrix elements appear.  These will be
given in units of $M^{}_{\text{{GT}}}$ 
and be denoted \cite{19} by $\chi$. First
those associated with the mass mechanism 
\begin{equation}
X_{\text{L}}^{} = \frac{<m_\nu>} {m_{\text{e}}} 
(\chi_{\text{F}}^{}-1) +
 <\frac{m_{\text{p}}} {M_{\text{N}}}>^{}_{\text{L}} 
\chi^{}_{\text{H}},
\label{eq:34}   
\end{equation}
\begin{equation}
X_{\text{R}} =
(\kappa^2 + \epsilon^2) <\frac{m_{\text{p}}} 
{M^{}_{\text{N}}}>^{}_{\text{R}} \chi^{}_{\text{H}},
\label{eq:35}   
\end{equation}
with 
\begin{equation}
\chi^{}_{\text{F}} = (\frac {g^{}_{\text{V}}}{g^{}_{\text{A}}})^2 
\frac {M_{\text{F}}}{M_{\text{GT}}},  
\label{eq:36}   
\end{equation}
\begin{equation}
\chi^{}_{\text{H}} 
= ((\frac {g^{}_{\text{V}}}{g^{}_{\text{A}}})^2  
M^{}_{\text{FH}} - M^{}_{\text{GTH}})/ M^{}_{\text{GT}}.  
\label{eq:37}   
\end{equation}
Secondly those not connected with the mass mechanism: 
$\chi^{}_{\text{F}\omega}$, $\chi_{\text{GT}\omega}$, 
$\chi^{}_{\text{R}}$, $\chi_{1^\pm}$, $\chi_{2^\pm}$
$\chi^{\prime}_{\text{F}}$, $\chi^{\prime}_{\text{GT}}$, 
$\chi^{\prime}_{\text{T}}$,  $\chi^{\prime}_{\text{P}}$
where
\begin{equation}
\chi^{}_{\text{F}\omega} =  (\frac {g^{}_{\text{V}}}
{g^{}_{\text{A}}})^2 \frac{M_{\text{F}\omega}}{M_{\text{GT}}},  
\label{eq:38}   
\end{equation}
\begin{equation}
\chi^{}_{\text{GT}\omega} =  
\frac{M_{\text{GT}\omega}}{M_{\text{GT}}},  
\label{eq:39}   
\end{equation}
\begin{equation}
\chi^{}_{\text{R}} =  \frac{M^{}_{\text{R}}}{M_{\text{GT}}}, 
\label{eq:40}   
\end{equation}
and
\begin{equation}
\chi_{1^\pm} = \pm 3\chi^\prime_{\text{F}} 
+ \chi^\prime_{\text{GT}} -6\chi^\prime_{\text{T}},
\label{eq:41}   
\end{equation}
\begin{equation}
\chi_{2^\pm} = \pm\chi_{\text{F}\omega} 
+ \chi_{\text{GT}\omega} - \frac {1} {9}
 \chi_{1^\pm} 
\label{eq:42}   
\end{equation}
in an obvious notation ($\chi^\prime_{\text{F}} = 
M^\prime_{\text{F}} /M_{\text{GT}}$ etc.).
In the limit in which the energy denominator 
$\epsilon_{\text{m}}$ can be neglected
we obtain 
\begin{equation}
\chi^{}_{\text{F}} = \chi^\prime_{\text{F}} = \chi_{\text{F}\omega}, 
\label{eq:43}   
\end{equation}
\begin{equation}
\chi_{\text{GT}} = \chi^\prime_{\text{GT}} = 
\chi_{\text{GT}\omega} = 1.
\label{eq:44}   
\end{equation}

As we have already mentioned the closure approximation was not employed in
our approximation.  In writing, however, the expression for 
$\epsilon_{\text{m}}$ we
made the standard approximation of replacing the electron energy by an
average value.  Furthermore the lepton wave functions were taken out of the 
nuclear integral incorporating their effect via a distortion factor (see
refs. \cite{18,19}).

In avoiding the closure approximation \cite{26,27,28,39} the momentum
space representation was found extremely useful.  
As it was shown in ref. \cite{39}
by exploiting the properties of the harmonic oscillator wave functions it
was possible to express the energy dependent radial integrals of each type
of operator in addition to the energy in terms of only two parameters $n$
and $\ell$.  The parameter $\ell$ takes values
 ($\ell=0,1,2$ for scalar, vector and tensor rank respectively).  
$n$ also takes
few values limited by the number ${\tilde N}$  of oscillator quanta involved
in the two particle wave function of the interacting nucleons 
($n \leq 2{\tilde N}$) regardless of the
number of single particle configurations employed \cite{39}.  
In the systems we studied $n \leq 10$.

\section{ The QRPA formalism with proton-neutron pairing}

The QRPA formalism employed in the $\beta\beta$-decay, a modification of the
usual RPA formalism that involves a change of charge by two units, is
fairly well known \cite{26}-\cite{37}.  
So we will briefly mention its main features
here.   $\beta\beta$ decay is viewed as a two step process.  In the first
step a proton particle neutron hole is created on the initial state
$|0^+_{\text{i}}>$ 
and associated with the intermediate states $|mJM>$ of the
$(A,Z+1)$ nuclear system.  In the second step, rather than considering an
additional proton particle - neutron hole acting on the excited
intermediate states, we consider a neutron particle-proton hole on the final
state  $|0^+_{\text{f}}>$  leading also to states of the
$(A,Z+1)$ nucleus labeled  $|{\overline m}JM>$.  
The states $|mJM>$ and $|\overline{ m}JM>$ are obtained
by solving the well known QRPA equations.  Unfortunately they
are not orthogonal to each other.  The nuclear physics dependence of
$\beta\beta$-decay is essentially contained in the matrix element
\begin{eqnarray}
\lefteqn{
<0^+_{\text{f}} \parallel {\tilde u}^{\text{k}} (a^+_{\text{pc}} 
a^{}_{\text{nd}}) 
\parallel {\overline m}J> }&& \nonumber \\
&&\times <{\overline m} J| m J>
<mJ \parallel u^{\text{k}} 
(a^+_{\text{pa}} a^{}_{\text{nb}}) \parallel 0^+_{\text{i}}>, 
\label{eq:45}    
\end{eqnarray}
where  $u^{\text{k}}_{\lambda} 
(a^+ a)$ are the usual tensors of rank k (k =
J here) built of protons and neutrons as indicated. 
${\tilde u}^{\text{k}}_{\lambda} 
(a^+ a)=(-1)^{\text{k}-\lambda}u^{\text{k}}_{-\lambda} (a^+ a)$ 
is the time reversed operator of $u^{\text{k}}_{\lambda} (a^+ a)$.
$a,b,c,d$ designate
all the single particle quantum numbers except for the angular momentum
projection $m_{\text{a}}$  i.e.,
$a \Longleftrightarrow 
(n_{\text{a}} \ell_{\text{a}} j_{\text{a}})$ etc. 
The overlap is necessary since, as we have
mentioned above, these intermediate states are not orthogonal to each
other.  Hence the energy denominator $\epsilon_{\text{m}}$ 
encountered in the previous section is given by the prescription
\begin{eqnarray}
\epsilon_{\text{m}} & = & 
E_{\text{m}} - <E_{\text{e}}> - E_{\text{f}} = 
E_{\text{m}} -  \frac {1} {2} 
(E_{\text{i}} +E_{\text{f}}) \nonumber \\
& \rightarrow & \frac {1} {2} (E_{\text{m}} +E_{\overline {\text{m}}})
- \frac {1} {2} (E_{\text{i}}+E_{\text{f}}) \nonumber \\
& = & \frac {1}{2} (\Omega_{\text{m}} + \Omega_{\overline {\text{m}}}).
\label{eq:46}
\end{eqnarray}
Here, $\Omega_{\text{m}}$ and 
$\Omega_{\overline {\text{m}}}$ are the QRPA energies of the
excited states $|m J M>$ and $|{\overline m} J M>$ 
calculated from the ground state energy of the initial 
and final nucleus, respectively.

The next step consists in expressing the tensor operators 
$u^{\text{k}}_{\lambda} (a^+ a)$ in terms of
quasiparticles \cite{26} i.e.,
\begin{equation} 
\left( \matrix{ a^+_{\text{pam}_{\text{a}}} \cr 
a^{}_{\text{pa}{\tilde {\text{m}}}_{\text{a}}} } \right) = 
\left( \matrix{ u_{\text{p}} (a) & -v_{\text{p}} 
(a) \cr v_{\text{p}} (a) & ~~u_{\text{p}} (a) } \right) 
\left( \matrix{ b^+_{1\text{am}_{a}} \cr 
 b^{}_{1\text{a}{\tilde {\text{m}}}_{\text{a}}} } \right),  
\label{eq:47}    
\end{equation}
\begin{equation} 
\left( \matrix{ a^+_{{\text{n a m}}_{\text{a}}} \cr 
 a^{}_{\text{na}{\tilde {\text{m}}}_{\text{a}}} } \right) = 
\left( \matrix{ u_{\text{n}} (a) & -v_{\text{n}} (a) \cr 
v_{\text{n}} (a) & ~~u_{\text{n}} (a) } \right) 
\left( \matrix{ b^+_{{\text{2am}}_{a}} \cr 
b^{}_{ {\text{2a}}{\tilde {\text{m}} }_{\text{a}} } } \right),
\label{eq:48}    
\end{equation} 
where $b^+$ and b are the construction and destruction operators for
quasiparticles. The tilde  $\sim$ indicates the time reversed states
$a_{\text{a}{\tilde {\text{m}}}_{\text{a}}}$ 
= $(-)^{{\text{j}}_{\text{a}}-{\text{m}}_{\text{a}}}$ 
$ a_{{\text{a-m}}_{\text{a}}}$  etc.    
Clearly since protons and neutrons do not mix the indices 1 and 2 can be
identified with protons and neutrons respectively.  The parameters $u$ and
$\upsilon$  are the occupation probabilities obtained by solving
the standard BCS equations for the initial state.  A similar set of
equations for the final states $(A,Z+2)$ yields the occupation
probabilities ${\overline u}$ and ${\overline v}$ 
entering the matrix element in the left of Eq.\ (\ref{eq:45}).

The QRPA states $|mJM>$ are of the form \cite{29,30,31}
\begin{eqnarray}
|mJM> &=& {Q^{\text{m}}_{\text{JM}}}^{+} 
|0^+_i>^{}_{\text{RPA}} \nonumber \\
& \equiv & \sum^{}_{\text{a, b}} 
~[ X^{\text{m}}_{12}(a,b,J)B^+_{1 2}(a,b,J,M)  \nonumber \\
&-&Y^{\text{m}}_{12}(a,b,J)
{\tilde B}^{}_{1 2}(a,b,J,M) ] |0^+_{\text{i}}>,
\label{eq:49}    
\end{eqnarray}
where
\begin{eqnarray}
\lefteqn{
B^+_{\mu \nu}(a, b, J, M) = n(\mu a, \nu b)}&& \nonumber \\
&& \times \sum^{}_{{\text{m}}_{\text{a}} , {\text{m}}_{\text{b}} }
C^{\text{JM}}_{{\text{j}}_{\text{a}}{\text{m}}_{\text{a}}
{\text{j}}_{\text{b}}{\text{m}}_{\text{b}} } 
b^+_{\mu{\text{am}}_{\text{a}}} b^+_{\nu{\text{bm}}_b}~ 
(\mu ,\nu = 1,2), 
\label{eq:50}    
\end{eqnarray}
\begin{equation}
{\tilde B}^{}_{\mu \nu}
(a, b, J M) = (-1)^{\text{J-M}} B^{}_{\mu \nu}(a, b, J, -M),
\label{eq:51}    
\end{equation}
and 
\begin{equation}
n(\mu a, \nu b) = \frac{(1+(-1)^{\text{J}} 
\delta_{\mu \nu} \delta_{\text{ab}})}
{(1+\delta_{\mu \nu} \delta_{\text{ab}})^{3/2}}.
\label{eq:52}    
\end{equation}
The forward- and backward- going amplitudes $X^{\text{m}}_{12}$ and
$Y^{\text{m}}_{12}$ and the energies of the excited states 
$\Omega_{\text{m}}$ are
obtained by solving the QRPA equation 
for the initial nucleus (A,Z) \cite{27}-\cite{31}.
By performing the QRPA diagonalization for the final nucleus (A,Z+2)
we obtain the amplitudes $X^{\overline {\text{m}}}_{12}$, 
$Y^{\overline {\text{m}}}_{12}$
and eigenenergies $\Omega_{\overline {\text{m}}}$ of the QRPA state 
$|{\overline m} JM>$.

One can show that  
\begin{eqnarray}
\lefteqn{
<m J \parallel u^{\text{k}}(a^+_{\text{pa}} 
a_{\text{nb}}) \parallel 0^+_{\text{i}}>} 
& & \nonumber \\ 
& & = \delta_{\text{kJ}}  ~\sqrt{2J+1}~
[u_{\text{p}}(a) \upsilon_{\text{n}}(b) 
X^{\text{m}}_{12}(a,b,J) \nonumber \\
& & \phantom{\delta_{kJ} ~\sqrt{2J+1}~ }
+ \upsilon_{\text{p}}(a) u_{\text{n}}(b) 
Y^{\text{m}}_{12}(a,b,J)],
\label{eq:53}
\end{eqnarray} 
\begin{eqnarray}
\lefteqn{<0^+_{\text{f}} \parallel 
{\tilde u}^k(a^+_{\text{pc}} a_{\text{nd}}) 
\parallel {\overline m} J> } & & \nonumber \\ 
& & = \delta_{\text{kJ}} ~ \sqrt{2J+1} ~
[ {\bar \upsilon}_{\text{p}}(c) {\bar u}_{\text{n}}(d)  
X^{\overline {\text{m}}}_{12}(c,d,J)
\nonumber \\ 
& & \phantom{\delta_{kJ} ~\sqrt{2J+1}~ }
+ {\bar u}_{\text{p}}(c) {\bar \upsilon}_{\text{n}}(d) 
Y^{{\overline {\text{m}}}}_{12}(c,d,J)].
\label{eq:54}
\end{eqnarray}
The overlap integral takes the form
\begin{eqnarray}
\lefteqn{
<{\overline m} J|m J> = \sum_{\text{a, b}}[ 
X^{\text{m}}_{12}(a,b,J) 
X^{{\overline {\text{m}}}}_{12}(a,b,J) } & & \nonumber \\
& & \phantom{<{\overline m} J|m J> = \sum_{a,b}[ }
- Y^{\text{m}}_{12}(a,b,J) Y^{{\overline {\text{m}}}}_{12}(a,b,J)].
\label{eq:55}     
\end{eqnarray}
Once proton-neutron correlations are introduced  the quasiparticle
labels can no longer be identified with protons and neutrons, but they
become  mere
labels.  The transformation matrix \cite{40}-\cite{46}
is generalized to the $4 \times 4$
matrix
\begin{eqnarray}
\lefteqn{
\left(\begin{array}{c}a^+_{{\text{pam}}_{\text{a}}}\\
a^+_{{\text{nam}}_{\text{a}}}\\
a_{{\text{pa}{\tilde {\text{m}}}}_{\text{a}}} \\
 a_{{\text{na}{\tilde {\text{m}}}}_{\text{a}}}\end{array}\right) 
 = \left(\begin{array}{cccc} u_{\text{1p}}(a) & 
          u_{\text{2p}}(a) & - \upsilon_{\text{1p}}(a) & 
- \upsilon_{\text{2p}}(a)\\
 u_{\text{1n}}(a) & u_{\text{2n}}(a) 
& - \upsilon_{\text{1n}}(a) & - \upsilon_{\text{2n}}(a)\\
 \upsilon_{\text{1p}}(a) &  \upsilon_{\text{2p}}(a) 
& ~u_{\text{1p}}(a) & ~u_{\text{2p}}(a)\\
 \upsilon_{\text{1n}}(a) &  \upsilon_{\text{2n}}(a) 
& ~u_{\text{1n}}(a) & ~u_{\text{2n}}(a) 
\end{array}\right)
 } & & \nonumber \\
&&
\phantom{ \begin{array}{cccc} u_{\text{1p}}(a) & 
          u_{2p}(a) & - \upsilon_{1p}(a) & 
- \upsilon_{2p}(a)\\
 u_{1n}(a) & u_{2n}(a) & - \upsilon_{1n}(a) & - \upsilon_{2n}(a)\\
 \upsilon_{1p}(a) &  \upsilon_{2p}(a) & ~u_{1p}(a) & ~u_{2p}(a)\\
 \upsilon_{1n}(a) &  \upsilon_{2n}(a) & ~u_{1n}(a) & ~u_{2n}(a) 
\end{array} }
\times \left(\begin{array}{c} b^+_{{\text{1am}}_{\text{a}}}\\
b^+_{{\text{2am}}^{}_{\text{a}} }\\
b_{{\text{1a}{\tilde {\text{m}}}}_{\text{a}}} \\
 b_{ {\text{2a}{\tilde {\text{m}}}}_{\text{a}} }\end{array}\right)
\label{eq:56}
\end{eqnarray}
independently of the angular momentum projection quantum number 
$m_{\text{a}}$.

The columns of the above matrix are the eigenvectors of the generalized
Hartree-Fock-Bogoliubov (HFB) equations (see ref. \cite{44,45,46} for
details).  As such are, of course, determined only up to an overall phase
(for each one).  

If the p-n pairing interaction is switched on, 
the angular momentum-coupled phonon operator takes the form \cite{44,45,46}
\begin{eqnarray}
Q^{\text{m+}}_{\text{JM}} & = &
~ \sum_{\text{a, b}} 
\{ X^{\text{m}}_{12}(a,b,J) B^{+}_{12}(a,b,J,M) \nonumber \\
&&~~~~~~~~~~~
+ Y^{\text{m}}_{12}(a,b,J) {\tilde B}_{12}(a,b,J,M) \}\nonumber \\
&+& \sum_{a \leq b \atop \mu = 1,2}
\{ X^{\text{m}}_{\mu \mu}(a,b,J)  B^{+}_{\mu \mu}(a,b,J,M) \nonumber \\
&&~~~~~~~~~~~
+ Y^{m}_{\mu \mu}(a,b,J) {\tilde B}^{}_{\mu \mu}(a,b,J,M) \},
\label{eq:57}    
\end{eqnarray}
The amplitudes $X^{\text{m}}$, $Y^{\text{m}}$ 
are obtained by solving the QRPA
eigenproblem
\begin{eqnarray}
\left( \matrix{ A & B \cr B & A } \right)
\left( \matrix{ X^{\text{m}} \cr 
Y^{\text{m}}} \right) = \Omega_{\text{m}} 
\left( \matrix{ 1 & 0 \cr 0 & -1 } \right)
\left( \matrix{ X^{\text{m}} \cr Y^{\text{m}} } \right),
\label{eq:58}    
\end{eqnarray}
with
\begin{eqnarray}
A =   
\left( \matrix{ A^{11, 11} & A^{11, 22} & A^{11, 12} \cr 
 A^{22, 11} & A^{22, 22} & A^{22, 12} \cr 
 A^{12, 11} & A^{12, 22} & A^{12, 12} } \right),
\label{eq:59}    
\end{eqnarray}
\begin{eqnarray}
 B = 
\left( \matrix{ B^{11, 11} & B^{11, 22} & B^{11, 12} \cr 
 B^{22, 11} & B^{22, 22} & B^{22, 12} \cr 
 B^{12, 11} & B^{12, 22} & B^{12, 12} } \right) ,
\label{eq:60}    
\end{eqnarray}
and
\begin{eqnarray}
X^{\text{m}} = \left( \matrix{ X^{\text{m}}_{11} \cr  
X^{\text{m}}_{22} \cr  X^{\text{m}}_{12} } \right), ~~  
Y^{\text{m}} = \left( \matrix{ Y^{\text{m}}_{11} \cr 
Y^{\text{m}}_{22} \cr Y^{\text{m}}_{12} } \right).
\label{eq:61}    
\end{eqnarray}
The QRPA equation in Eq.\ (\ref{eq:58}) 
represents a general equation for all 
excited states of a given even-even nucleus. The explicit form of the 
A and B submatrices has been reviewed in our previous publications
\cite{44,45,46}. 
We note that if proton-neutron pairing interaction is neglected,
the QRPA submatrices $ A^{1 2, 1 1}$, $ A^{1 1, 1 2}$, 
$ A^{1 2, 2 2}$ $ A^{2 2, 1 2}$ and 
$ B^{1 2, 1 1}$, $ B^{1 1, 1 2}$,
$ B^{1 2, 2 2}$, $ B^{2 2, 1 2}$  are equal to zero and the 
QRPA equation  posesses two types of eigenstates: The eigenstates 
of the first type ( I )
with wave functions 
$X^{\text{m}}=(X^{\text{m}}_{11},X^{\text{m}}_{22},0)$, 
$Y^{\text{m}}=(Y^{\text{m}}_{11},Y^{\text{m}}_{22},0)$ 
having the origin in proton-proton and neutron-neutron quasiparticle
excitations. The eigenstate of the second type ( II ) with the wave 
functions $X^{\text{m}}=(0,0,X^{\text{m}}_{12})$, 
$Y^{\text{m}}=(0,0,Y^{\text{m}}_{12})$ 
generated by the phonon operator in Eq.\ (\ref{eq:49}).

Eqs. (\ref{eq:53}), and (\ref{eq:54}) are generalized as follows
\begin{eqnarray}
\lefteqn{<m J \parallel u^{\text{k}}
(a^+_{\text{pa}} a_{\text{nb}}) \parallel 0^+_{\text{i}}> = 
\delta_{\text{kJ}}\sqrt{2J+1} } & &\nonumber \\
&& \times \sum_{\mu,\nu = 1,2} m(\mu a, \nu b) 
[u_{\mu p}(a) \upsilon_{\nu n}(b) 
X^{\text{m}}_{\mu \nu}(a,b,J) \nonumber \\
& & \phantom{ \sum_{\mu,\nu = 1,2} }~~~~~
+\upsilon _{\mu \text{p}}(a) u_{\nu \text{n}}(b) 
Y^{\text{m}}_{\mu \nu}(a,b,J)],
\label{eq:62}    
\end{eqnarray}
\begin{eqnarray}
\lefteqn{<0^+_{\text{f}} \parallel {\tilde u}^{\text{k}}
(a^+_{\text{pc}} a_{\text{nd}}) 
\parallel {\overline m} J>  = \delta_{\text{kJ}}
\sqrt{2J+1} } & & \nonumber \\
&& \times \sum_{\mu,\nu = 1,2} m(\mu a, \nu b) 
[{\bar \upsilon}_{\mu{\text{p}}}(c) {\bar u}_{\nu{\text{n}}}(d) 
X^{\overline {\text{m}}}_{\mu \nu}(c,d,J) \nonumber \\
& & \phantom{ \sum_{\mu,\nu = 1,2} }~~~~~
+{\bar u}_{\mu{\text{p}}}(c) {\bar \upsilon}_{\nu{\text{n}}}(d) 
Y^{\overline {\text{m}}}_{\mu \nu}(c,d,J)],
\label{eq:63}    
\end{eqnarray}
with 
\begin{equation}
m(\mu a, \nu b) = {(1+(-1)^J \delta_{\mu \nu} \delta_{ab})}/
{(1+\delta_{\mu \nu} \delta_{a b})^{1/2}}.
\label{eq:64}    
\end{equation} 
We note that the $X^{\text{m}}_{\mu \nu}(a,b,J)$ 
and $Y^{\text{m}}_{\mu \nu}(a,b,J)$
amplitudes are calculated by the QRPA equation in Eq.\ (\ref{eq:58})
only for the configurations $\mu a \leq \nu b$ ( i.e.,
$\mu = \nu$ and the orbitals are ordered $a \leq b$ and 
$\mu = 1$, $\nu = 2$ and the orbitals are not ordered).
For different configurations 
$X^{\text{m}}_{\mu \nu}(a,b,J)$ and $Y^{\text{m}}_{\mu \nu}(a,b,J)$ 
in Eqs.\ (\ref{eq:62}), and (\ref{eq:63})
are given as follows:
\begin{equation}
X^{\text{m}}_{\mu \nu}(a,b,J)
 = -(-1)^{{\text{j}}_{\text{a}}+{\text{j}}_{\text{b}}-\text{J}}
X^{\text{m}}_{\nu \mu}(b,a,J), 
\label{eq:65}    
\end{equation} 
\begin{equation}
Y^{\text{m}}_{\mu \nu}(a,b,J) =
 -(-1)^{{\text{j}}_{\text{a}}+{\text{j}}_{\text{b}}-J}
Y^{\text{m}}_{\nu \mu}(b,a,J). 
\label{eq:66}    
\end{equation} 
Note that in the limit in which there is no
proton-neutron pairing, i.e., $u_{\text{2p}} = \upsilon_{\text{2p}} 
= u_{\text{1n}} = \upsilon_{\text{1n}} = 0$, 
Eqs.\ (\ref{eq:62}), and (\ref{eq:63})  reduce
to Eqs.\ (\ref{eq:53}), and (\ref{eq:54})
 respectively by setting  $u_{\text{1p}} = u_{\text{p}},  
\upsilon_{\text{1p}} =
\upsilon_{\text{p}}, u_{\text{2n}} = u_{\text{n}}$, 
and $\upsilon_{\text{2n}} = \upsilon_{\text{n}}$. Clearly,
in the case without proton-neutron pairing the eigenstates of type I 
do not contribute to the beta decay transition matrix elements
in Eqs.\ (\ref{eq:62}), and  (\ref{eq:63}). 

The overlap integral becomes
\begin{eqnarray}
\lefteqn{
<{\overline m} J|m J> = \sum_{\mu a \leq \nu b}[ 
X^{\text{m}}_{\mu \nu}(a,b,J) 
X^{{\overline {\text{m}}}}_{\mu \nu}(a,b,J) } 
& & \nonumber \\
& & \phantom{<{\overline m} J|m J> = \sum_{a,b}[ }
- Y^{\text{m}}_{\mu \nu}(a,b,J) 
Y^{{\overline {\text{m}}}}_{\mu \nu}(a,b,J)].
\label{eq:67}     
\end{eqnarray}
By setting $X^{\text{m}}_{\mu \mu} = Y^{\text{m}}_{\mu \mu} =
X^{\overline {\text{m}}}_{\mu \mu} = 
Y^{\overline {\text{m}}}_{\mu \mu} = 0$  
the  above overlap is reduced to that of Eq.\ (\ref{eq:55}).

To complete the discussion we mention that the single particle wave
functions and energies were  obtained by using a Coulomb corrected Woods Saxon
potential.  The interaction employed was the Brueckner G-matrix which is a
solution of the Bethe-Goldstone equation employed using the Bonn 
OBEP \cite{47}. 
Proton and neutron number conservation in the initial and final state
was respected on the average with
\begin{equation}
(N_{\text{np}} -<N_{\text{np}}>)/ N_{\text{np}} \leq 10^{-4}. 
\label{eq:68}    
\end{equation}
The BCS $\text{pp}$ and 
$\text{nn}$ parameters  $d_{\text{pp}}$ and  $d_{\text{nn}}$ 
were obtained by fits to the
experimental proton and neutron gaps as in ref. \cite{44,45,46}.  
The $\text{np}$ strength parameter $d_{\text{np}}$ 
is fixed by a renormalization of the 
$T = 1 \  J = 0$ pairing force as in ref. \cite{44,45,46}.
In the QRPA calculations it is necessary to introduce renormalization
parameters $g_{\text{pp}}$ and $g_{\text{ph}}$ for 
the particle-particle and
particle-hole interactions, which in principle should be close to unity.  
Our adopted values were $g_{\text{pp}} = 1.0$ 
and $g_{\text{ph}} = 0.8$. 
For higher value  of $g_{\text{ph}}$ the particle-hole interaction for 
some multipolarities is too strong. The lowest eigenvalue becomes 
imaginary and leads to a collapse of the correlated ground state. 
By the method outlined above we obtained the matrix elements 
$M_{\text{GT}}^{0\nu}$, 
$\chi_{\text{F}}$, $\chi_{\text{H}}$, 
$\chi^\prime_{\text{F}}$, $\chi^\prime_{\text{GT}}$, 
$\chi^\prime_{\text{T}}$, $\chi^{}_{\text{F}\omega}$,
$\chi^{}_{\text{GT}\omega}$, $\chi^{\prime}_{\text{P}}$ 
and $\chi^{}_{\text{R}}$ for the nuclei  
$^{48}{\text{Ca}}$, $^{76}{\text{Ge}}$, 
$^{82}{\text{Se}}$, $^{96}{\text{Zr}}$, 
$^{100}{\text{Mo}}$, $^{116}{\text{Cd}}$,
$^{128}{\text{Te}}$, $ ^{130}{\text{Te}}$ 
and $^{136}Xe$ for the orbitals shown in 
Table \ref{table1}. 
These matrix elements are given in Table \ref{table2}.  
For comparison we also
present in the same Table \ref{table2} 
the values obtained without p-n pairing.  These
last results differ slightly from those of our earlier calculations 
\cite{26,27,28} due
to the different model space employed.  

By glancing at the Table \ref{table2}
 we see that the effect of the inclusion of the
p-n pairing is significant. 
Perhaps the most important
matrix element is  $|M^{0\nu}_{\text{GT}} (1-\chi_{\text{F}})|$, 
which connect us directly with the
effective neutrino mass $|<m_{\nu }>|$. We see that the inclusion of 
the p-n pairing reduces the value of this matrix element. The largest 
reductions of $|M^{0\nu}_{\text{GT}} (1-\chi_{\text{F}})|$  
by factors $30.2$,
$6.7$, $3.4$, $2.8$ and $2.3$ are associated with  $A = 100$,
 $96$, $48$, $128$ and $76$ systems.  By these factors 
also the lower limits on  $|<m_{\nu }>|$ are enhanced 
in respect to the calculations without p-n pairing.  To study the 
influence of p-n pairing on the evaluation of the limits on
lepton number non-conserving parameters of right-handed currents 
it is necessary to calculate the integrated kinematical factors 
$G^{}_{\text{0k}}$ \cite{19,20}. They are listed in Table \ref{table3}.  
A small differences with the values of ref. \cite{19} has origin  
in different adopted value of nuclear radius $R^{}_{0}$
[see Eq.\ (\ref{eq:18})]. The problem of the extraction of the
lepton number non-conserving parameters we shall study in Sec. IV.

In the nuclear systems 
$A = 96, 100$ and $116$ we have noticed a sensitivity of
this matrix element with 
respect to the number of orbitals employed.  Since
including in our present calculation with p-n 
pairing all the 15 orbitals
employed in the earlier calculations was prohibitive in term of 
computer time, we
decided to employ 12 orbitals.  This is admissible 
since we are interested in
compairing the results with and without p-n pairing 
in the same model space. 
Furthermore our present results without 
p-n pairing for $^{100}Mo$ agree with
those of ref. \cite{2} which used the same model space. 
The results so obtained for this nucleus  are comparable with
those of the other nuclei, which makes our choice reasonable.  
The other matrix
elements depend a bit more strongly on the p-n pairing correlations.  
The effect is even stronger for some individual multipoles, 
especially if the corresponding matrix element is suppressed. 

\section{Decay Rates}

The  $0\nu \beta \beta$-decay can be expressed in terms of the lepton
violating parameter $<m_\nu>/m_{\text{e}}$ etc. defined in Sec. II as 
follows \cite{27}, namely
\begin{eqnarray}
\lefteqn{
[T_{1/2}^{0\nu}]^{-1} = G^{0\nu}_{01} |M^{0\nu}_{\text{GT}}|^2 
\left\{ |X_{\text{L}}|^2 + |X_{\text{R}}|^2 - 
{\tilde C}_1^\prime X_{\text{L}} X_{\text{R}}
\right.  } & & \nonumber \\ 
&&+ {\tilde C}_2 |\lambda| X_{\text{L}} cos \psi_1 
 + {\tilde C}_3 |\eta| X_{\text{L}} cos \psi_2 
+ {\tilde C}_4 |\lambda|^2 + {\tilde C}_5 |\eta|^2 
\nonumber \\
&& \left. + {\tilde C}_6 
|\lambda||\eta| cos (\psi_1 -\psi_2) + Re ({\tilde C}_2 
\lambda X_{\text{R}} + {\tilde C}_3 \eta  X_{\text{R}}) \right\},
\label{eq:69}
\end{eqnarray}    
where $X_{\text{L}}$ and $X_{\text{R}}$ 
are defined in Eqs.\ (\ref{eq:34}), and 
(\ref{eq:35}). 
$\psi_1$ and $\psi_2$ are the relative phases between $X_{\text{L}}$ and
$\lambda$ and $X_{\text{L}}$ and $\eta$ respectively.  
The ellipses \{...\} indicate
contributions arising from other particles, e.g., intermediate SUSY particles 
\cite{48} or unusual particles which are predicted by superstring models
\cite{49} or exotic Higgs scalars \cite{50} etc.

The quantities $G^{0\nu}_{01}$ are calculated using the prescription of ref.
\cite{4}.  The coefficients  $C^\prime_1, C_{\text{i}}, 
i= 2-6$ are combinations
of kinematical functions and the nuclear matrix elements discussed in
the previous section.  They are defined as follows.
\begin{eqnarray}
{\tilde C}_2 &=& -(1 - \chi^{}_{\text{F}})
(\chi^{}_{2^-} {\tilde G}_{03} - \chi^{}_{1^+} {
\tilde G}_{04}), \nonumber \\
{\tilde C}_3 &=& - (1 - \chi^{}_{\text{F}}) 
(\chi^{}_{2^+} {\tilde G}_{03} - \chi^{}_{1^-}
{\tilde G}_{04} - \chi^{\prime}_{\text{P}} {\tilde G}_{05} + 
\chi^{}_{\text{R}} {\tilde G}_{06}),  \nonumber \\
{\tilde C}_4 &=& \chi_{2^-}^2 {\tilde G}_{02}+ \frac {1}{9}
\chi_{1^+}^2  
{\tilde G}_{04} -\frac {2}{9} \chi_{1^+}  \chi_{2^-} {\tilde
G}_{03},  \nonumber \\
{\tilde C}_5 &=& \chi_{2^+}^2   {\tilde G}_{02} + \frac
{1}{9} \chi_{1^-}^2  {\tilde G}_{04} 
-\frac {2}{9} \chi^{}_{1^-}  \chi^{}_{2^+} {\tilde G}_{03} 
\nonumber \\ 
&&~~~~~~~~~~~~~~~~ + (\chi_{\text{P}}^\prime)^2 {\tilde G}_{08} 
- \chi_{\text{P}}^\prime 
\chi^{}_{\text{R}} {\tilde G}_{07} 
+ \chi_{\text{R}}^2 {\tilde G}_{09}),  
\nonumber \\
{\tilde C}_6 &=& -2[ \chi^{}_{2^-} \chi^{}_{2^+}  {\tilde G}_{02} 
- \frac {1}{9} (\chi^{}_{1^+} \chi^{}_{2^+}~+~ 
\chi^{}_{2^-} \chi^{}_{1^-}) {\tilde G}_{03}
\nonumber \\
&&~~~~~~~~~~~~~~~~~~~~~~~~~~~~~~~~~
+  \frac {1}{9} \chi^{}_{1^+} \chi^{}_{1^-} {\tilde G}_{04}].
 \label{eq:70}
\end{eqnarray}
Here ${\tilde C}_1^\prime \cong 10(\epsilon_0^2 + 6\epsilon_0 +6)/
(\epsilon_0^4 
+10\epsilon_0^3 +10\epsilon_0^2 +60\epsilon_0 + 30)$,
$\epsilon_0$ is the available energy in electron mass units. 
$C_1^\prime$ is less than 10\% and it can be safely neglected. 

The quantities ${\tilde G}_{\text{0i}}$ are defined as follows: 
\begin{eqnarray}
{\tilde G}_{\text{0i}} &=& G_{\text{0i}}/G_{01}~~ (\text{i}=2,3,4), 
\nonumber \\
{\tilde G}_{05} &=& 2G_{05}/G_{01}, \nonumber \\
{\tilde G}_{06} &=& \frac {1}{4} m_{\text{e}} 
R_0 G_{06}/G_{01}, \nonumber \\
{\tilde G}_{07} &=& 2(\frac {1}{4} m_{\text{e}} R_0) G_{07}/G_{01},
\nonumber \\
{\tilde G}_{08} &=& 4G_{08}/G_{01}, \nonumber \\
{\tilde G}_{09} &=& (\frac {1}{4} m_{\text{e}} R_0)^2 G_{09}/G_{01}.
\label{eq:71}
\end{eqnarray}
The values of the parameters ${\tilde G}_{\text{0i}}, i = 2,...,6$ are
presented in Table \ref{table4}.  
The coefficients  ${\tilde C}_{\text{i}}, \text{i} = 2,...,6$ with
and without p-n pairing are shown in Table \ref{table5}.

The most stringent experimental limits are:
\begin{eqnarray}
A &=& 48: ~~ T_{1/2}^{0\nu} \geq 9.5 \times 10^{21} y  
~(76\% CL) ~ [5], \nonumber \\ 
A &=& 76: ~~ T_{1/2}^{0\nu} \geq 5.6 \times 10^{24} y 
~(90\% CL) ~ [2], \nonumber  \\
A &=& 82: ~~ T_{1/2}^{0\nu} \geq 2.7 \times 10^{22} y  
~(68\% CL) ~ [3], \nonumber  \\ 
A &=& 96: ~~ T_{1/2}^{\text{all}} \geq 3.9 \times 10^{19} y  
~ (geochem.) ~ [11], \nonumber  \\ 
A &=& 100: ~ T_{1/2}^{0\nu} \geq 4.4 \times 10^{22} y 
~(68\% CL) ~[4], \nonumber  \\
A &=& 116: ~ T_{1/2}^{0\nu} \geq 2.9 \times 10^{22} y 
~(90\% CL) ~[6], \nonumber  \\
A &=& 128: ~ T_{1/2}^{\text{all}} \geq 7.7 \times 10^{24} y  
~(geochem.) ~[7] \nonumber  \\
A &=& 130: ~ T_{1/2}^{0\nu} \geq 2.3 \times 10^{22} y 
~ (68\% CL) ~[8], \nonumber  \\
A &=& 136: ~ T_{1/2}^{0\nu} \geq 3.4 \times 10^{23} y 
~ (68\% CL) ~[9].  \nonumber
\end{eqnarray}

Using Eq.\ (\ref{eq:69}), the functions ${\tilde G}_{01}$ of 
Table \ref{table4} and 
the matrix elements $M^{0\nu}_{\text{GT}}$ of 
Table \ref{table2} with p-n pairing we get the
constrains:
\begin{eqnarray}
A &=& 48:  \ \ \  \{...\} \leq 8.0 \times 10^{-9}, \nonumber \\
A &=& 76:  \ \ \  \{...\} \leq 6.6 \times 10^{-12}, \nonumber \\
A &=& 82:  \ \ \  \{...\} \leq 7.9 \times 10^{-10}, \nonumber \\
A &=& 96:  \ \ \  \{...\} \leq 4.4 \times 10^{-6}, \nonumber \\
A &=& 100:  \ \   \{...\} \leq 1.2 \times 10^{-9}, \nonumber \\
A &=& 116:  \ \   \{...\} \leq 3.9 \times 10^{-8}, \nonumber \\
A &=& 128:  \ \   \{...\} \leq 3.8 \times 10^{-11}, \nonumber \\
A &=& 130:  \ \   \{...\} \leq 2.3 \times 10^{-10}, \nonumber \\
A &=& 136:  \ \   \{...\} \leq 1.4 \times 10^{-11}. \nonumber
\end{eqnarray}
Here \{...\} indicates the quantity which is enclosed in the curly braket of
Eq.\ (\ref{eq:69}).  Clearly the nucleus with the smallest value of \{...\}
is going to provide the most stringent limit on the lepton violating
parameters.  The life time itself is not a clear indicator since the
function $G_{01}$ varies from nucleus to nucleus.  Large $G_{01}$, i.e., large
phase space, leads to short life times for a given lepton violation parameter.
Thus the most stringent limits are expected from the A=76 system.

To impose limits on $X_{\text{L}}, X_{\text{R}}, 
\lambda, \eta$ one must make four -
dimensional plots making some assumptions about $\psi_1, \psi_2$ and the
relative signs of $\lambda$ and $\eta$ with $\chi_R$.  Then for a given
value of $X_{\text{L}}$ one can extract limits on $<m_\nu>$ 
+$m^{}_{\text{e}}
<\frac{ m_{\text{p}}}{M_{\text{N}}}>_{\text{L}}^{} 
{{\chi}^{}_{\text{H}}}/
{({\chi}^{}_{\text{F}}-1)}$. 
To extract limits on $<m_\nu>$ and  
$<\frac{1}{M_{\text{N}}}>$ 
one must make further plots (knowing $\chi^{}_{\text{H}}$,
$\chi^{}_{\text{F}}$ 
from the calculations). This is really a complicated 
procedure  to be worth doing only if and when $0\nu \beta\beta$-decay
is definitely seen.  At present we will constrain the above parameters by
assuming that one mechanism at a time dominates. 
The limits thus obtained appear
in Table \ref{table6}.  We must mention that in the case of heavy neutrino 
only the parameters ($<\frac{1}{M_{\text{N}}}>_{\text{R}} 
(\epsilon^2 + \kappa^2))^{-1}
= <\frac{1}{M_{\text{N}}}>^{-1}_{\text{L}}$ 
can be extracted this way.  The parameter
$<\frac{1}{M_{\text{N}}}>^{-1}_{\text{R}}$ shown in 
Table \ref{table6} was obtained by taking \cite{18}
$\epsilon^2 + \kappa^2 = 10^{-2}$. In line with what we mentioned
above  the extraction of  the parameter $\eta$ depends on ${\tilde C}_5$ alone.
In all cases ${\tilde C}_5$ is dominated by  
$\chi^{\prime}_{\text{P}}$ and /or
$\chi^{}_{\text{R}}$.  So in Table \ref{table7} 
we present two values of $\eta$, one with nuclear
recoil included and one without recoil.  With the possible exception of the
$A = 100$ and $A = 128, 130$ system, the recoil contribution is dominant. 
In the $\text{Te}$ isotope $\chi_{\text{P}}^\prime$ 
and $\chi^{}_{\text{R}}$ compete with each other.

Another lepton violating process is the $0\nu \beta\beta$- decay with
Majoron emission.  The corresponding expression is  
\begin{equation}
{\left( T_{1/2}^{0\nu,\chi^0} \right) }^{-1} = {|\eta_{\chi^0}|^2} 
G^{}_{0\nu, \chi^0} |M^{0\nu}_{\text{GT}} (1-\chi_{\text{F}})|^2,     
\label{eq:72}    
\end{equation}
where
\begin{eqnarray}
G_{0\nu, \chi^0} &=& {\tilde G}_{0 \chi^0} G_{01},\nonumber \\
{\tilde G}_{0 \chi^0} &=& \frac {1} {(2\pi)^2} \epsilon_0^2 \frac
{g_1(\epsilon_0)} {g_0(\epsilon_0)}, \nonumber \\
g_0(\epsilon_0) &=&  \epsilon_0^4 + 10 \epsilon_0^3 + 40 
\epsilon_0^2 +60 \epsilon_0 + 30,  \nonumber \\
g_1(\epsilon_0) &=&  \epsilon_0^4 + 14 \epsilon_0^3 + 84 
\epsilon_0^2 +210 \epsilon_0 + 210,  \nonumber \\
\eta_{\chi^0} &=& \sum_{\text{i,j}} U^{(11)}_{\text{ei}} 
U^{(11)}_{\text{ej}}\frac
{1}{\sqrt{2}} e^{\text{i}(\alpha_{\text{i}}-\alpha_{\text{j}})} 
g_{\text{ij}}.     
\label{eq:73}    
\end{eqnarray} 
$\epsilon^{}_{0}$ is the available energy  
in units of the electron mass. $g_{\text{ij}}$ is the 
coupling of the Majoron to the neutrino mass eigenstates, i.e., 
\begin{equation}
{\cal L} = \frac {g_{\text{ij}}}{\sqrt{2}} 
{\bar \nu}^{}_{\text{iL}} \nu^{}_{\text{jR}} \chi^0 + \text{h.c.}.
\label{eq:74}    
\end{equation}
It can also be written as
\begin{equation}
\frac {g_{\text{ee}}} {{\sqrt 2}} 
\cong\eta_{\chi^0}, \ \ \ {\cal L} = 
\frac {g_{\text{ee}}}{\sqrt{2}} 
\nu_{\text{eL}} \nu_{\text{eR}}^c \chi^0 + \text{h.c.}.     
\label{eq:75}    
\end{equation}
 The corresponding experimental limits are 
\begin{eqnarray}
A &=& 48:  ~~ T_{1/2}^{0\nu,\chi^0} > 7.2 \times 10^{20}
~ (90\% CL)~[14], \nonumber  \\ 
A &=& 76:  ~~T_{1/2}^{0\nu,\chi^0} > 3.9 \times 10^{22}
~(90\% CL)~[2], \nonumber  \\
A &=& 82:  ~~T_{1/2}^{0\nu, \chi^0} > 1.6 \times 10^{21}  
~(68\% CL) ~~[3], \nonumber \\ 
A &=& 96:  ~~T_{1/2}^{\text{all}} > 3.9 \times 10^{19} 
~(geochem.) ~[11], \nonumber \\ 
A &=& 100:  ~T_{1/2}^{0\nu, \chi^0} > 7.9 \times 10^{20} 
~(68\% CL) ~[15], \nonumber  \\
A &=& 116:  ~T_{1/2}^{0\nu, \chi^0} > 1.8 \times 10^{19} 
~(99\% CL) ~[6], \nonumber  \\
A &=& 128:  ~T_{1/2}^{\text{all}} > 7.7 \times 10^{24}  
~(geochem.) ~[7], \nonumber  \\
A &=& 130:  ~T_{1/2}^{\text{all}} > 2.7 \times 10^{21}  
~(geochem.) ~[7], \nonumber  \\
A &=& 136:  ~T_{1/2}^{0\nu, \chi^0} > 4.9 \times 10^{21} 
~(90\% CL) ~[9]. \nonumber
\end{eqnarray}
From the above experimental limits and the values of 
$G_{01, \chi^0}$ (Table \ref{table4}) 
and $|M^{0\nu}_{\text{GT}}(1-\chi_{\text{F}})|$ 
(Table \ref{table2}) we obtain the limits of 
$|\eta_{\chi^0}|$ listed below:
\begin{eqnarray}
A &=& ~48:  \ \ \  |\eta_{\chi^0}| < 7.0  \times 10^{-4}, 
\nonumber \\
A &=& ~76:  \ \ \  |\eta_{\chi^0}| < 1.4  \times 10^{-4}, 
\nonumber \\
A &=& ~82:  \ \ \  |\eta_{\chi^0}| < 1.9  \times 10^{-4}, 
\nonumber \\
A &=& ~96:  \ \ \  |\eta_{\chi^0}| < 3.6  \times 10^{-3}, 
\nonumber \\
A &=& 100:  \ \ \  |\eta_{\chi^0}| < 9.9  \times 10^{-3}, 
\nonumber \\
A &=& 116:  \ \ \  |\eta_{\chi^0}| < 2.6  \times 10^{-3}, 
\nonumber \\
A &=& 128:  \ \ \  |\eta_{\chi^0}| < 5.7  \times 10^{-5}, 
\nonumber \\
A &=& 130:  \ \ \  |\eta_{\chi^0}| < 1.5  \times 10^{-4}, 
\nonumber \\
A &=& 136:  \ \ \  |\eta_{\chi^0}| < 1.3  \times 10^{-4}. 
\nonumber 
\end{eqnarray}

\section{ Summary and conclusions}

In the present work we have evaluated the nuclear matrix elements 
entering the double beta decay of the experimentally most
interesting nuclear systems. We have employed the Quasiparticle 
Random Phase Approximation which seems to be the most practical 
method of nuclear structure calculation of nuclear systems 
which are far away from closed shells. In these calculations we have
included the proton-neutron pairing correlations which have been
neglected in the previous calculations. We have found that such 
correlations have important effects on all the needed matrix 
elements and should not be neglected. The magnitude of the effect
depends, of course, on the type of operator employed, i.e., on the
mechanism for the $0\nu\beta\beta$-decay. We will concentrate our
discussion on those matrix elements which were not unusually 
suppressed. We will give their value both without p-n pairing
correlations and after such correlations are turned on. Since we have 
assumed that one mechanism at a time is important  we can summarize our 
results as follows. 

i) Light neutrino mass mechanism. The relevant nuclear matrix element
is $|M^{0\nu}_{\text{GT}}(1-{\chi}^{}_{\text{F}})|^2$. 
The five largest matrix 
elements are $9.2$, $6.2$, $5.8$, $5.6$ and $4.9$ associated with
$A=76, 128, 96, 130$ and $82$ respectively. Once p-n pairing correlations
are turned on they become $1.8$, $0.77$, $0.13$, $2.2$ and $2.7$
respectively, i.e., we have a reduction factor ranging between $2$
and $8$. In the most interesting case of the $A=76$ system we have 
a reduction of about $5$. The effect on the extraction on the neutrino
mass is less pronounced. (The square root of the above factor).

ii) Heavy intermediate neutrino mass mechanism. The relevant nuclear
matrix element is $|{\chi}^{}_{\text{H}} M^{0\nu}_{\text{GT}}|^2$. 
The five largest 
matrix elements are $4.0\times 10^4$,  $1.1\times 10^4$, $1.0\times 10^4$,    
$0.98\times 10^4$ and $0.96\times 10^4$ for     
$A=76, 48, 128, 96$ and $82$ respectively. With p-n pairing 
correlations they become $0.37\times 10^4$, $6.0$, $0.19\times 10^4$,
$1.4\times 10^2$ and $0.43\times 10^4$. Notice the almost 
complete suppression for $A = 48$ and $96$ and the large reduction
factor for $A = 76$ (about 10). 

iii) The mass independent $\lambda$- mechanism. The relevant matrix
element is ${\tilde C}^{}_{4}|M^{0\nu}_{\text{GT}}|^2$. 
The largest matrix 
elements  $14.5$, $14.1$, $13.6$, $13.4$ and $6.2$ are associated with
$A = 96$, $76$, $82$, $130$ and $136$ respectively. They become $0.49$, 
$3.07$, $6.5$, $8.1$ and $3.8$ respectively. Notice that quite 
unexpectedly the matrix element of the $A = 128$ system is much 
smaller than that for the $A = 130 $ system. In the case of the 
$A = 76 $  we have a reduction factor of about  $5$.

iv) The neutrino mass independent $\eta$- mechanism. The relevant matrix
element is ${\tilde C}^{}_{5}|M^{0\nu}_{\text{GT}}|^2$. 
The largest values are
$4.1\times 10^5$, $3.7\times 10^5$, $2.9\times 10^5$,
$1.4\times 10^5$ and $9.1\times 10^4$ associated with $A = 76, 128, 130,$
$136$ and $96$. They are reduced to $5.5\times 10^4$, $2.4\times 10^5$, 
$1.7\times 10^5$, $1.8\times 10^5$ and $1.7\times 10^4$. In the case
of the $A = 76$ system the reduction factor is about $7$.

We do not fully understand why the effect of the p-n pairing 
correlations should be so large. In the case of $A = 100$  this effect  
is much more dramatic.  On the contrary for the $A = 136$ system
the matrix elements are 
fairly large and not much effected by such correlations. Finally 
we do not understand why the effect of p-n correlations is so 
different on the two Tellurium isotopes. From Eq.\ (\ref{eq:69})
it is clear that the $A = 76$ system provides the most stringent
limits on the lepton violating parameters. This is partly due 
to the large matrix elements obtained for this system but mainly 
due to the fact that the experimental life-time limit is the best. 
Unfortunately,  the introduction of p-n pairing correlations makes 
the extracted limits  of the lepton violating parameters 
less stringent. In  fact we find 
\begin{eqnarray}
|<m_{\nu }>| &<& \left\{
\begin{array}{c} 0.8~ eV \\ 1.8~ eV \end{array} \right., \nonumber \\
|<\frac{1}{M_{\text{N}}}>^{}_{\text{L}}|^{-1} &>& \left\{
\begin{array}{c} 3.8\times 10^7 ~GeV \\ 
1.6\times 10^7 ~GeV \end{array} \right., \nonumber \\
|<\frac{1}{M_{\text{N}}}>^{}_{\text{R}}|^{-1} &>& \left\{ 
\begin{array}{c} 3.8\times 10^5 ~GeV \\ 
1.6\times 10^5 ~GeV \end{array} \right., \nonumber \\
| <\lambda> | &<& \left\{
\begin{array}{c} 1.3\times 10^{-6}  \\ 
2.7\times 10^{-6}  \end{array} \right., \nonumber \\
| <\eta> | &<& \left\{ 
\begin{array}{c} 7.4\times 10^{-9}  \\ 
2.0\times 10^{-8}  \end{array} \right., \nonumber \\
|\eta_{\chi^0}| &<& \left\{ 
\begin{array}{c} 8.4\times 10^{-5}  \\ 
1.9\times 10^{-4}  \end{array} \right.. \nonumber
\end{eqnarray}
In the above expressions the upper (lower) values correspond to the case
without (with) p-n pairing correlations. 
It is interesting to note that
 the $A = 128$ system, in spite of the 
fact that its $0\nu \beta\beta$ the decay width is kinematically
suppressed, provides 
quite stringent limits on the lepton violating parameters with 
the possible exception of $\lambda $. This is quite surprising 
since the nuclear matrix elements involved are not favored compared 
to the $A = 76$ system and in any case they should not be very 
different from those of the $A = 130$ system. Furthermore the extracted limits 
on the lepton violating parameters will become even more stringent if 
the $0\nu $ life time is used since 
$T_{1/2}^{0\nu} \geq T_{1/2}^{{all}}$. We are, therefore, 
inclined to suspect that the life time of $T_{1/2}^{0\nu} $
 $ \geq 5.6\times 10^{24} $ years is quite a bit exaggerated. We will 
not, however, elaborate further on this  controversial point
concerning the life time of this long lived isotope. 

Finally we should mention that the above extracted limits still
suffer from uncertainties of nuclear origin. We should not 
forget that the effect of the interference between the various
mechanisms has not been taken into account.

\onecolumn
\widetext

\begin{table}[t]
\caption{The single particle orbitals used in the present calculations.}
\label{table1}
\begin{tabular}{cccccccccc}
 & & & & & & & & & \\ 
nuclear & $^{48}Ca$ & $^{76}Ge$ & $^{82}Se$ & $^{96}Zr$ 
& $^{100}Mo$ & $^{116}Cd$ & $^{128}Te$ & $^{130}Te$ & $^{136}Xe$ \\ 
orbitals & & & & & & & & & \\ 
 & & & & & & & & & \\ \hline
 & & & & & & & & & \\
1 & $ 0 s_{1/2}$ & $ 1 s_{1/2}$ & $ 1 s_{1/2}$ & $ 0f_{7/2}$ & $ 0f_{7/2}$ &
$ 0f_{5/2}$  & $ 1p_{3/2}$ & $ 1p_{3/2}$ & $ 1p_{3/2}$ \\
2 & $ 0p_{3/2}$ & $ 0f_{7/2}$ & $  0f_{7/2}$ & $0f_{5/2}$ & $0f_{5/2}$ &
$1p_{3/2}$ & $ 1p_{1/2}$ & $ 1p_{1/2}$ & $ 1p_{1/2}$ \\
3 & $ 0p_{1/2}$ & $ 0f_{5/2}$ & $  0f_{5/2}$ & $1p_{3/2}$ & $1p_{3/2}$ &
$1p_{1/2}$ & $ 0g_{7/2}$ & $  0g_{7/2}$ & $  0g_{7/2}$ \\
4 & $ 0d_{5/2}$ & $ 1p_{3/2}$ & $  1p_{3/2}$ & $1p_{1/2}$ & $1p_{1/2}$ &
$0g_{9/2}$ & $ 0g_{7/2}$ & $  0g_{7/2}$ & $  0g_{7/2}$ \\
5 & $ 0d_{3/2}$ & $ 1p_{1/2}$ & $  1p_{1/2}$ & $0g_{9/2}$ & $0g_{9/2}$ &
$0g_{7/2}$ & $ 1d_{5/2}$ & $ 1d_{5/2}$ & $ 1d_{5/2}$ \\
6 & $ 1s_{1/2}$ & $0g_{9/2}$ & $0g_{9/2}$ & $0g_{7/2}$ & $0g_{7/2}$ &
$1d_{5/2}$ & $ 1d_{3/2}$ & $ 1d_{3/2}$ & $ 1d_{3/2}$ \\
7 & $ 0f_{7/2}$ & $0g_{7/2}$ & $0g_{7/2}$ & $1d_{5/2}$ & $1d_{5/2}$ &
$1d_{3/2}$ & $ 2s_{1/2}$ & $ 2s_{1/2}$ & $ 2s_{1/2}$ \\
8 & $ 0f_{5/2}$ & $1d_{5/2}$ & $1d_{5/2}$ & $1d_{3/2}$ & $1d_{3/2}$ &
$2s_{1/2}$ & $0h_{11/2}$ & $ 0h_{11/2}$ & $0h_{11/2}$ \\
9 & $1p_{3/2}$ & $1d_{3/2}$ & $1d_{3/2}$ & $2s_{1/2}$ & $2s_{1/2}$ &
$0h_{11/2}$ & $0h_{9/2}$ & $ 0h_{9/2}$ & $0h_{9/2}$ \\
10 & $1p_{1/2}$ & $2s_{1/2}$ & $2s_{1/2}$ & $0h_{11/2}$ & $0h_{11/2}$ &
$0h_{9/2}$ & $1f_{7/2}$ & $ 1f_{7/2}$ & $1f_{7/2}$ \\
11 &  &  &  & $0h_{9/2}$ & $0h_{9/2}$ & $1f_{7/2}$  & & &  \\
12 &  &  &  &  & & $1f_{5/2}$  & & &  \\
\end{tabular}
\end{table}

\widetext
\begin{table}[h]
\caption{The matrix elements of ${0\nu\beta\beta}$-decay  
for $^{48}{\text{Ca}}$,
$^{76}{\text{Ge}}$, $^{82}{\text{Se}}$, $^{96}{\text{Zr}}$, 
$^{100}{\text{Mo}}$, $^{116}{\text{Cd}}$, $^{128}{\text{Te}}$,
$^{130}{\text{Te}}$ and $^{136}{\text{Xe}}$ 
calculated in the framework of QRPA with and
without p-n pairing.}
\label{table2}
\begin{tabular}{cccccccccc} 
 & & & & & & & & & \\ 
nucleus & $^{48}Ca$ & $^{76}Ge$ & $^{82}Se$ & $^{96}Zr$ & $^{100}Mo$ &
 $^{116}Cd$ & $^{128}Te$ & $^{130}Te$ & $^{136}Xe$ \\ 
 & & & & & & & & & \\ \hline
 & & & & & & & & & \\
\multicolumn{10}{c}{ QRPA without p-n pairing} \\
 & & & & & & & & & \\ 
 $ M^{0\nu}_{\text{GT}}$ 
& -0.785 & 2.929 & -2.212 &  2.097 & 0.615 & 0.449 &
 2.437 & 2.327 &  1.598 \\
 ${\chi}^{}_{\text{F}}$ &   -0.468 &   -0.038 & -0.008 &  
-0.149 & -0.766 &   -1.103 & -0.0179 &  -0.004 &  0.028 \\
$|M^{0\nu}_{\text{GT}} (1-\chi_{\text{F}})|$ 
& 1.152 & 3.040 & 2.230 & 2.409 & 1.086 &  
 0.944 & 2.480 & 2.335 & 1.553 \\
 ${\chi}^{}_{\text{H}}$ 
& -134.9 & -68.37 & -44.27 & -47.24 & -124.8 & -47.06 
& -41.54 & -39.82 & -21.92 \\
 & & & & & & & & & \\
 ${\acute{\chi}}^{}_{\text{F}}$ &  
-0.504 & -0.035 & -0.004 & -0.168 & -0.817 & 
 -1.173 & -0.022 &  -0.007 & 0.022 \\
 ${\acute{\chi}}^{}_{\text{GT}}$ 
& 0.975 & 1.077 &  1.050 & 1.143 & 1.174 & 1.074 &
 1.097 & 1.097 &  1.123 \\
 ${\acute{\chi}}^{}_{\text{T}}$ 
& -0.212 &  0.244 & 0.079 &  0.121 & -0.477 & -0.812 &
 0.307 &  0.282 & 0.349 \\
 ${\chi}^{}_{\text{F}\omega}$ 
& -0.437 & -0.038 & -0.013 & -0.130 & -0.709 &
 -1.032 & -0.012 &  0.001 &  0.036 \\
 ${\chi}^{}_{\text{GT}\omega}$ 
& 1.057 & 0.916 & 0.960 &  0.845 &  0.683 &
 0.859 &  0.894 & 0.895 & 0.875 \\
 ${\acute{\chi}}^{}_{\text{P}}$ 
&  0.168 & -1.147 & -0.049 &  -0.836 & -3.843 &
 -3.891 & -1.400 & -1.451 & -1.627 \\
 & & & & & & & & & \\
 $\chi^{}_{\text{R}}$ 
& 172.1 &  193.0 & 124.2 & 113.8 & 105.1 & -151.5 & 157.1 &
 149.0 & 124.8 \\
 & & & & & & & & & \\
\multicolumn{10}{c}{ QRPA with p-n pairing} \\
 & & & & & & & & & \\ 
 $M^{0\nu}_{\text{GT}}$ 
& -0.405 & 1.846 & -1.153 & 0.280 & -0.584 & 0.119 &
 1.270 &   1.833 &   1.346 \\
 $\chi_{\text{F}}$ 
& 0.158 & 0.274 & -0.416 &  2.282 & 0.939 & -6.784 &
 0.308 & 0.184 & 0.066 \\
 $|M^{0\nu}_{\text{GT}} (1-\chi_{\text{F}})|$ 
& 0.341 & 1.340 & 1.633 & 0.358 & 
 0.036 & 0.926 & 0.879 & 1.495 & 1.257 \\
 $\chi_{\text{H}}$ 
& 6.075 & -32.75 & -57.20 & -41.64 & -14.22 & -453.8 &
 -34.02 & -55.72 & -35.37 \\
 & & & & & & & & & \\
 ${\acute{\chi}}_{\text{F}}$ 
& 0.184 & 0.322 & -0.467 & 2.601 & 1.067 &
 -7.400 & 0.370 & 0.218 & 0.082 \\
 ${\acute{\chi}}_{\text{GT}}$ 
& 1.226 & 1.124 & 1.082 & 1.587 &  0.934 &
 0.927 &  1.159 &  1.115 &  1.167 \\
 ${\acute{\chi}}_{\text{T}}$ 
& 0.130 & 0.214 &  0.179 &  0.209 &  0.853 &
   -3.991 &  0.343 & 0.411 & 0.332 \\
 ${\chi}_{\text{F}\omega}$ 
& 0.131 & 0.235 &  -0.379 & 2.069 &   
 0.812 & -6.170 &  0.260 & 0.159 &  0.052 \\
 ${\chi}_{\text{GT}\omega}$ 
& 0.775 & 0.876 & 0.927 & 0.335 &  1.142 &
 0.938 & 0.831 & 0.879 & 0.832 \\
 ${\acute{\chi}}_{\text{P}}$ 
& -0.009 & -0.479 &  -1.621 & -4.802 & 2.519 &
 -7.592 & -2.907 & -0.993 & -2.441 \\
 & & & & & & & & & \\
 $\chi_{\text{R}}$ 
&  57.32 & 129.3 & 131.1 & 157.3 & 162.2 & -333.6 & 158.6 &
 192.6 & 138.4 \\
\end{tabular}
\end{table}

\newpage
\widetext
\begin{table}[t]
\caption{The integrated kinematical factors  
$G_{0k}$ for $0^{+}\rightarrow{0^{+}}$
transition of $(\beta\beta)_{0\nu}-decay$. The definition 
of $G_{0k}$ is given in ref. [19] in Eqs.\ (3.5.17-21).}
\label{table3}
\begin{tabular}{cccccccccc} 
 & & & & & & & & & \\
\multicolumn{10}{c}{ $(\beta\beta)_{0\nu}-decay: 
0^{+}\rightarrow{0^{+}}$ 
 transition} \\
 & & & & & & & & & \\ \cline{2-10}
 & & & & & & & & & \\ 
 & $^{48}Ca$ & $^{76}Ge$ & $^{82}Se$ & $^{96}Zr$ & $^{100}Mo$ &
 $^{116}Cd$ & $^{128}Te$ & $^{130}Te$ & $^{136}Xe$ \\ 
 & & & & & & & & & \\ \hline
 $(E_{\text{i}}-E_{\text{f}})$ [MeV] 
& 5.294 & 3.067 & 4.027 & 4.372 & 4.055 & 
 3.830 & 1.891 & 3.555 & 3.503 \\
 & & & & & & & & & \\ 
  $G_{01}$\hspace{1mm} $[10^{-14}y^{-1}]$ &  8.031 &  0.7928 &
  3.524 & 7.362 & 5.731 &  6.233 & 
2.207$\times10^{-1}$ &  5.543 &  5.914 \\
  $G_{02}$\hspace{1mm} $[10^{-13}y^{-1}]$ &  5.235 &   0.1296 &
  1.221 & 3.173 & 2.056 &  1.957 & 
6.309$\times10^{-3}$ &  1.441 &  1.483  \\
  $G_{03}$\hspace{1mm} $[10^{-14}y^{-1}]$ &  6.037 &   0.4376 &  
  2.413 & 5.380 & 4.036 &  4.305 & 
6.177$\times10^{-2}$ &  3.669 &  3.890  \\
  $G_{04}$\hspace{1mm} $[10^{-14}y^{-1}]$ &  1.705 &   0.1538 &  
  0.724 &  1.530 &  1.178 &  1.269 & 
3.368$\times10^{-2}$ & 1.113 & 1.183  \\
  $G_{05}$\hspace{1mm} $[10^{-12}y^{-1}]$ &  1.265 &  0.253 &    
  0.931 &  2.009 &  1.718 &  2.118 & 
1.390$\times10^{-1}$ & 2.083 &  2.298   \\
  $G_{06}$\hspace{1mm} $[10^{-11}y^{-1}]$ &  1.398  &  0.196 &       
  0.665 &  1.226 &  1.009 &  1.103 & 
6.969$\times10^{-2}$ & 1.011 &  1.077  \\
  $G_{07}$\hspace{1mm} $[10^{-10}y^{-1}]$ &  11.46 &  1.248 &         
  5.523 & 12.07 &  9.563 &  10.69 & 
4.363$\times10^{-1}$ & 9.544 &  10.25  \\
  $G_{08}$\hspace{1mm} $[10^{-11}y^{-1}]$ &  5.247 &  0.793 &           
  3.852 &  9.886 &  8.109 &  10.20 & 
4.227$\times10^{-1}$ & 9.749 &  10.84 \\
  $G_{09}$\hspace{1mm} $[10^{-9 }y^{-1}]$ &  6.262 &   0.491 &           
  1.980 &  3.686 &  2.819 & 2.800 & 
1.125$\times10^{-1}$ & 2.335 &  2.424 \\
 & & & & & & & & & \\ 
  $G_{01,{\chi}^{0}_{}}$\hspace{1mm} 
$[10^{-14}y^{-1}]$ &  2.425 &  0.0763 &
  0.6202 & 1.5315 & 1.0230 &  0.9879 & 
5.206$\times10^{-3}$ &  0.7487 & 0.7734 
 \\ 
\end{tabular}
\end{table}

\widetext
\begin{table}[h]
\caption{The kinematical functions ${\tilde{G}}_{0i}$, 
$i=2-9$. They are given in the notation of Pantis et al. [27].}
\label{table4}
\begin{tabular}{ccccccccc} 
 & & & & & & & &  \\
nuclear &  ${\tilde{G}}_{02}$ & ${\tilde{G}}_{03}$  &
${\tilde{G}}_{04}$ & ${\tilde{G}}_{05}$ & ${\tilde{G}}_{06}$  &
${\tilde{G}}_{07}$ & ${\tilde{G}}_{08}$ & ${\tilde{G}}_{09}$ \\  
transition & & & & & & & & \\
 & & &  & & & & & \\ \hline
$^{48}Ca\rightarrow{^{48}Ti}$ &  
6.518 & 0.752 & 0.212 &   31.50 &  0.450 &  73.87 & 2613. 
& 0.522 \\
$^{76}Ge\rightarrow{^{76}Se}$ &  
1.635 & 0.552 & 0.194 &  63.93 &  0.745 &  95.01 & 4001. & 
0.563 \\
$^{82}Se\rightarrow{^{82}Kr}$ &  
3.465 & 0.685 & 0.205 &   52.9 &  0.584 &  96.99 & 4372. & 
0.538 \\
$^{96}Zr\rightarrow{^{96}Mo}$ &  
4.310 & 0.731 & 0.208 &  54.58 &  0.543 &  107.0 & 5371. & 
0.5324 \\
$^{100}Mo\rightarrow{^{100}Ru}$ & 
3.588 & 0.704 & 0.206 &  59.97 &  0.582 &   110.3 &  5660.  
& 0.5375 \\
$^{116}Cd\rightarrow{^{116}Sn}$ & 
3.140 & 0.691 & 0.204 &  67.94 &  0.614 &   119.1 &  6547.  
& 0.5419 \\
$^{128}Te\rightarrow{^{128}Xe}$ & 
0.286 & 0.280 & 0.153 &  126.0 &  1.133 &   141.9 &  7662.  
& 0.6565 \\
$^{130}Te\rightarrow{^{130}Xe}$ & 
2.599 & 0.662 & 0.201 &  75.15 &  0.658 &   124.2 &  7035.  
&  0.5483 \\
$^{136}Xe\rightarrow{^{136}Ba}$ & 
2.507 & 0.658 & 0.200 &  77.72 &  0.667 &   127.0 &  7331.  
&  0.5497 
\\ 
\end{tabular}
\end{table}

\newpage
\widetext
\begin{table}[t]
\caption{The coefficients ${\tilde{C}}_{\text{i}}, 
\text{i}=1,2,3,4,5,6$ which are
combinations of nuclear matrix elements and phase space 
factors [see Eq.\ (55) of the text]
needed for the extraction of $|<m_{\nu}>|$, $\lambda$, 
$\eta$ etc. from the data.}
\label{table5}
\begin{tabular}{ccccccc}
 & & & & & &  \\
nuclear & ${\tilde{C}}_{1}$ &  ${\tilde{C}}_{2}$ & 
${\tilde{C}}_{3}$  &
${\tilde{C}}_{4}$ & ${\tilde{C}}_{5}$ & ${\tilde{C}}_{6}$  \\
transition & & & & & &  \\
 & & & & & & \\ \hline
 & & & & & &  \\
\multicolumn{7}{c}{ QRPA without p-n pairing} \\
 & & & & & &  \\ 
$^{48}Ca\rightarrow{^{48}Ti}$ &
 2.154 & -0.96 & $ -1.05\times{10^2}$ & 7.43 & 
$1.34\times{10^4}$ & -6.92 \\
$^{76}Ge\rightarrow{^{76}Se}$ &  
 1.078 & -0.66 & $ -2.26\times{10^2}$ & 1.65 & 
$4.73\times{10^4}$ & -3.10 \\
$^{82}Se\rightarrow{^{82}Kr}$ &  
 1.015 & -0.51 & $ -7.61\times{10}$ & 2.78 & 
$8.90\times{10^3}$ & -5.42 \\
$^{96}Zr\rightarrow{^{96}Mo}$ &  
 1.321 & -0.75 & $ -1.23\times{10^2}$ & 3.29 & 
$2.08\times{10^4}$ & -5.32 \\
$^{100}Mo\rightarrow{^{100}Ru}$ & 
 3.118 & -0.26 & $ -5.12\times{10^2}$   & 1.51 & 
$1.34\times{10^5}$ & 1.13 \\
$^{116}Cd\rightarrow{^{116}Sn}$ &
 4.421 & -0.18 & $ -3.55\times{10^2}$ & 2.03 & 
$4.13\times{10^4}$ &  2.35 \\
$^{128}Te\rightarrow{^{128}Xe}$ & 
 1.036 & -0.40 & $ -3.61\times{10^2}$ & 0.33 & 
$6.24\times{10^4}$ &  -0.65 \\
$^{130}Te\rightarrow{^{130}Xe}$ & 
 1.007 & -0.76 & $ -2.08\times{10^2}$ & 2.48 & 
$ 5.38\times{10^4}$ & -4.98 \\
$^{136}Xe\rightarrow{^{136}Ba}$ & 
 0.944 & -0.78 & $ -2.04\times{10^2}$ & 2.43 & 
$ 5.37\times{10^4}$ & -5.16 \\
 & & & & & &  \\
\multicolumn{7}{c}{ QRPA with p-n pairing} \\
 & & & & & &  \\ 
$^{48}Ca\rightarrow{^{48}Ti}$ &
 0.708 & -0.24 & $ -2.25\times{10} $ &  2.72 & 
$1.76\times{10^3}$ & -6.67 \\
$^{76}Ge\rightarrow{^{76}Se}$ &  
 0.526 & -0.19 & $ -9.27\times{10}$ & 0.90 & 
$1.62\times{10^4}$ & -2.53 \\
$^{82}Se\rightarrow{^{82}Kr}$ &  
 2.005 & -1.52 & $ -2.30\times{10^2}$ & 4.86 & 
$4.13\times{10^4}$ & -5.41 \\
$^{96}Zr\rightarrow{^{96}Mo}$ &  
 1.643 & -3.01 & $ ~4.48\times{10^2}$ &  6.24 & 
$2.18\times{10^5}$ & 17.6 \\
$^{100}Mo\rightarrow{^{100}Ru}$ & 
 0.0037 & -0.062 & 3.27 &  4.95 & 
$4.99\times{10^3}$ & -19.0 \\
$^{116}Cd\rightarrow{^{116}Sn}$ &
 60.59 &  -5.86 & $ -2.35\times{10^3}$ & 10.4 & 
$ 1.36\times{10^5}$ & 70.8 \\
$^{128}Te\rightarrow{^{128}Xe}$ & 
 0.479 &  -0.13 & $ -3.78\times{10^2}$ & 0.17 & 
$1.47\times{10^5}$ & -0.56 \\
$^{130}Te\rightarrow{^{130}Xe}$ &
 0.665 & -0.62 & $ -1.65\times{10^2}$ &  2.41 & 
$5.11\times{10^4}$ & -5.91 \\
$^{136}Xe\rightarrow{^{136}Ba}$ & 
 0.872 & -0.66 & $-2.64\times{10^2}$ &  2.11 & 
$ 9.71\times{10^4}$ & -4.53 \\
\\ 
\end{tabular}
\end{table}

\widetext
\begin{table}[h]
\caption{ The limits on lepton number non-conserving parameters
$|<m_{\nu}>|$, $|<\frac{1}{M_{\text{N}}>^{}_{\text{L}}}|^{-1},$  and 
$|<\frac{1}{M_{\text{N}}>^{}_{\text{R}}}|^{-1}$ 
deduced from the experimental limits
of $0\nu\beta\beta$ decay lifetimes for the nuclei 
studied in this work. For the extraction of the parameter 
$|<\frac{1}{M_{\text{N}}}>^{}_{\text{R}}|^{-1}$ the
value of $\varepsilon^{2}+\kappa^{2}=10^{-2}$ was assumed. 
Only one term was assumed dominant. 
$T^{0\nu\text{-1eV}}_{1/2}$ is the calculated
 $0\nu\beta\beta$ half-life times assuming 
$|<m_{\nu}>|$ =\text{ 1eV}.  }
\label{table6}
\begin{tabular}{cccccc} 
 & & & & &  \\
nucleus & $|<m_{\nu }>|$ & $|<\frac{1}{M_{N}}>^{}_{L}|^{-1} $  & 
 $ {|<\frac{1}{M_{N}}>^{}_{R}|^{-1}}$  & 
 $T^{0\nu\text{-1eV}}_{1/2}$ & $T^{0\nu\text{-exp}}_{1/2}$ \\
 & & & & &  \\
 & [eV] & [GeV] & [GeV] & [years]  &  [years ] ref.  \\
 & & & & &  \\ \hline
 & & & & & \\
\multicolumn{6}{c}{ QRPA without p-n pairing} \\
 & & & & &  \\ 
$^{48}Ca$ &
$\le 16$ & $\ge 1.9\times{10^{6}}$ & $\ge 1.9\times{10^{4}}$ &  
$2.5\times{10^{24}}$ &
$\ge 9.5\times{10^{21}}$ \cite{5} \\
$^{76}Ge$ &  
$\le 0.8$ & $\ge 3.8\times{10^{7}}$ & $\ge 3.8\times{10^{5}}$ & 
$3.6\times{10^{24}}$ &
$\ge 5.6\times{10^{24}}$ \cite{2} \\
$^{82}Se$ &  
$\le 7.4$ & $\ge 2.8\times{10^{6}}$ & $\ge 2.8\times{10^{4}}$ &  
$1.5\times{10^{24}}$ &
$\ge 2.7\times{10^{22}}$ \cite{3} \\
$^{96}Zr$ &  
$\le 125$ & $\ge 1.4\times{10^{5}}$ & $\ge 1.4\times{10^{3}}$ & 
$6.1\times{10^{23}}$ &
$\ge 3.9\times{10^{19}}$ \cite{11} \\
$^{100}Mo$ & 
$\le 9.4$ & $\ge 2.0\times{10^{6}}$ & $\ge 2.0\times{10^{4}}$ & 
$3.9\times{10^{24}}$ &
$\ge 4.4\times{10^{22}}$ \cite{4} \\
$^{116}Cd$ &
$\le 13$  & $\ge 4.0\times{10^{5}}$ & $\ge 4.0\times{10^{3}}$ & 
$4.7\times{10^{24}}$ &
$\ge 2.9\times{10^{22}}$ \cite{6} \\
$^{128}Te$ & 
$\le 1.6$ & $\ge 1.2\times{10^{7}}$ & $\ge 1.2\times{10^{5}}$ & 
$1.9\times{10^{25}}$ &
$\ge 7.3\times{10^{24}}$ \cite{7} \\
$^{130}Te$ & 
$\le 6.1$ & $\ge 3.1\times{10^{6}}$ & $\ge 3.1\times{10^{4}}$ & 
$8.6\times{10^{23}}$ &
$\ge 2.3\times{10^{22}}$ \cite{10} \\
$^{136}Xe$ & 
$\le 1.7$ & $\ge 6.6\times{10^{6}}$ & $\ge 6.6\times{10^{4}}$ & 
$1.8\times{10^{24}}$ &
$\ge 6.4\times{10^{23}}$ \cite{9} \\
 & & & & & \\
\multicolumn{6}{c}{ QRPA with p-n pairing} \\
 & & & & & \\ 
$^{48}Ca$ &
$\le 54$  & $\ge 7.6\times{10^{4}}$ & $\ge 7.6\times{10^{2}}$ & 
$2.8\times{10^{25}}$ &
$\ge 9.5\times{10^{21}}$ \cite{5} \\
$^{76}Ge$ &  
$\le 1.8$ & $\ge 1.6\times{10^{7}}$ & $\ge 1.6\times{10^{5}}$ & 
$1.8\times{10^{25}}$ &
$\ge 5.6\times{10^{24}}$ \cite{2} \\
$^{82}Se$ &  
$\le 10$  & $\ge 1.3\times{10^{6}}$ & $\ge 1.3\times{10^{4}}$ & 
$2.8\times{10^{24}}$ &
$\ge 2.7\times{10^{22}}$ \cite{3} \\
$^{96}Zr$ &  
$\le 841$ & $\ge 1.4\times{10^{4}}$ & $\ge 1.4\times{10^{2}}$ & 
$2.7\times{10^{25}}$ &
$\ge 3.9\times{10^{19}}$ \cite{11} \\
$^{100}Mo$ & 
$\le 285$  & $\ge 6.4\times{10^{6}}$ & $\ge 6.4\times{10^{4}}$ & 
$3.6\times{10^{27}}$ &
$\ge 4.4\times{10^{22}}$ \cite{4} \\
$^{116}Cd$ &
$\le 13$ &  $\ge 2.8\times{10^{5}}$ & $\ge 2.8\times{10^{3}}$ & 
$4.9\times{10^{24}}$ &
$\ge 2.9\times{10^{22}}$ \cite{6} \\
$^{128}Te$ & 
$\le 4.6$ & $\ge 7.4\times{10^{6}}$ & $\ge 7.4\times{10^{4}}$ & 
$1.5\times{10^{26}}$ &
$\ge 7.3\times{10^{24}}$ \cite{7} \\
$^{130}Te$ &
$\le 9.6$ & $\ge 4.2\times{10^{6}}$ & $\ge 4.2\times{10^{4}}$ & 
$2.1\times{10^{24}}$ &
$\ge 2.3\times{10^{22}}$ \cite{10} \\
$^{136}Xe$ & 
$\le 2.1$ & $\ge 9.3\times{10^{6}}$ & $\ge 9.3\times{10^{4}}$ & 
$2.8\times{10^{24}}$ &
$\ge 6.4\times{10^{23}}$ \cite{9} 
\\ 
\end{tabular}
\end{table}

\widetext
\begin{table}[t]
\caption{ The limits on lepton number non-conserving parameters
 $|<\lambda>|$ and $|<\eta>|$ deduced from the experimental limits
of $0\nu\beta\beta$ decay lifetimes for the nuclei studied in this 
work. For $<\eta>$ our results with and 
without inclusion of the recoil 
nuclear matrix elements are presented.
Only one term was assumed dominant. }
\label{table7}
\begin{tabular}{ccccc} 
 & & & &   \\
nucleus & 
$ | <\lambda> |$ & \multicolumn{2}{c}{$ | <\eta> |$} & 
$T^{0\nu\text{-exp}}_{1/2}(0\nu)$ \\
 & & & & \\
 & & without & with &  [years] ref. \\
 & & recoil & recoil & \\
 & & & &   \\ \hline
 & & & &  \\
 & \multicolumn{3}{c}{ QRPA without p-n pairing} \\
 & & & &   \\ 
$^{48}Ca$ &
 $\le 1.7\times{10^{-5}}$ &   $\le 5.3\times{10^{-6}}$ &
$\le 4.0\times{10^{-7}}$ &   
$\ge 9.5\times{10^{21}}$ \cite{5} \\
$^{76}Ge$ &  
 $\le 1.3\times{10^{-6}}$ &  $\le 2.2\times{10^{-8}}$ &
$\le 7.4\times{10^{-9}}$ &   
$\ge 5.6\times{10^{24}}$ \cite{2} \\
$^{82}Se$ &  
 $\le 8.8\times{10^{-6}}$ &  $\le 4.1\times{10^{-6}}$ &
$\le 1.5\times{10^{-7}}$ &   
$\ge 2.7\times{10^{22}}$ \cite{3} \\
$^{96}Zr$ &  
 $\le 1.6\times{10^{-4}}$ &  $\le 4.6\times{10^{-6}}$ &
$\le 1.9\times{10^{-6}}$ &   
$\ge 3.9\times{10^{19}}$ \cite{11} \\
$^{100}Mo$ & 
$\le 2.6\times{10^{-5}}$ &  $\le 1.1\times{10^{-7}}$ &
$\le 8.8\times{10^{-8}}$ &   
$\ge 4.4\times{10^{22}}$ \cite{4} \\
$^{116}Cd$ &
 $\le 3.7\times{10^{-5}}$ &  $\le 1.7\times{10^{-7}}$ &
$\le 2.6\times{10^{-7}}$ &   
$\ge 2.9\times{10^{22}}$ \cite{6} \\
$^{128}Te$ & 
 $\le 5.6\times{10^{-6}}$ &  $\le 2.6\times{10^{-8}}$ &
$\le 1.3\times{10^{-8}}$ &   
$\ge 7.3\times{10^{24}}$ \cite{7} \\
$^{130}Te$ & 
 $\le 7.6\times{10^{-6}}$ &  $\le 9.9\times{10^{-8}}$ &
$\le 5.2\times{10^{-8}}$ &   
$\ge 2.3\times{10^{22}}$ \cite{10} \\
$^{136}Xe$ & 
 $\le 2.1\times{10^{-6}}$ &  $\le 2.3\times{10^{-8}}$ &
$\le 1.4\times{10^{-8}}$ &   
$\ge 6.4\times{10^{23}}$ \cite{9} \\
 & & & & \\
 & \multicolumn{3}{c}{ QRPA with p-n pairing} \\
 & & & & \\ 
$^{48}Ca$ &
 $\le 5.4\times{10^{-5}}$ &  $\le 4.3\times{10^{-5}}$ &
$\le 2.1\times{10^{-6}}$ & 
$\ge 9.5\times{10^{21}}$ \cite{5} \\
$^{76}Ge$ &  
 $\le 2.7\times{10^{-6}}$ &  $\le 8.5\times{10^{-8}}$ &
$\le 2.0\times{10^{-8}}$ & 
$\ge 5.6\times{10^{24}}$ \cite{2} \\
$^{82}Se$ &  
 $\le 1.3\times{10^{-5}}$ &  $\le 2.6\times{10^{-7}}$ &
$\le 1.4\times{10^{-7}}$ & 
$\ge 2.7\times{10^{22}}$ \cite{3} \\
$^{96}Zr$ &  
 $\le 8.4\times{10^{-4}}$ &  $\le 6.0\times{10^{-6}}$ &
$\le 4.5\times{10^{-6}}$ & 
$\ge 3.9\times{10^{19}}$ \cite{11} \\
$^{100}Mo$ & 
 $\le 1.5\times{10^{-5}}$ &  $\le 1.8\times{10^{-7}}$ &
$\le 4.8\times{10^{-7}}$ & 
$\ge 4.4\times{10^{22}}$ \cite{4} \\
$^{116}Cd$ &
 $\le 6.1\times{10^{-5}}$ &  $\le 3.2\times{10^{-7}}$ &
$\le 5.4\times{10^{-7}}$ & 
$\ge 2.9\times{10^{22}}$ \cite{6} \\
$^{128}Te$ & 
 $\le 1.5\times{10^{-5}}$ &  $\le 2.4\times{10^{-8}}$ &
$\le 1.6\times{10^{-8}}$ & 
$\ge 7.3\times{10^{24}}$ \cite{7} \\
$^{130}Te$ &
 $\le 9.8\times{10^{-6}}$ &  $\le 1.8\times{10^{-7}}$ &
$\le 6.8\times{10^{-8}}$ & 
$\ge 2.3\times{10^{22}}$ \cite{10} \\
$^{136}Xe$ & 
 $\le 2.6\times{10^{-6}}$ &  $\le 1.8\times{10^{-8}}$ &
$\le 1.2\times{10^{-8}}$ & 
$\ge 6.4\times{10^{23}}$ \cite{9} 
\\ 
\end{tabular}
\end{table}

\end{document}